\documentclass[preprint,aps,tightenlines,showpacs,amsmath,amssymb,pra,superscriptaddress]{revtex4-1}
\usepackage{graphicx}
\usepackage{color}
\usepackage[nooneline,FIGTOPCAP,raggedright]{subfigure}
\usepackage{esint}
\usepackage{xparse}
\usepackage{soul}

\makeatletter
 \@ifundefined{textcolor}{}
 {%
   \definecolor{BLACK}{gray}{0}
   \definecolor{WHITE}{gray}{1}
   \definecolor{RED}{rgb}{1,0,0}
   \definecolor{GREEN}{rgb}{0,1,0}
   \definecolor{BLUE}{rgb}{0,0,1}
   \definecolor{CYAN}{cmyk}{1,0,0,0}
   \definecolor{MAGENTA}{cmyk}{0,1,0,0}
   \definecolor{YELLOW}{cmyk}{0,0,1,0}
 }

\NewDocumentCommand{\sotwo}{O{red}O{black}+m}
    {%
        \begingroup
        \setulcolor{#1}%
        \setul{-.5ex}{.4pt}%
        \def\SOUL@uleverysyllable{%
            \rlap{%
                \color{#2}\the\SOUL@syllable
                \SOUL@setkern\SOUL@charkern}%
            \SOUL@ulunderline{%
                \phantom{\the\SOUL@syllable}}%
        }%
        \ul{#3}%
        \endgroup
    }
\makeatother

\begin{document}
\title{Optimal control for fast and high-fidelity quantum gates in
  coupled superconducting flux qubits}

\author{Shang-Yu Huang}
%\author{Chia-Hsien Huang}
\author{Hsi-Sheng Goan}
\email{goan@phys.ntu.edu.tw}
\affiliation{Department of Physics and Center for Theoretical Sciences,
National Taiwan University, Taipei 10617, Taiwan}
\affiliation{Center for Quantum Science and Engineering, and National Center for Theoretical Sciences,
National Taiwan University, Taipei 10617, Taiwan}
%^{1,2}$\footnote{Author to whom any correspondence should be addressed.}}

\date{\today}

\begin{abstract}
We apply the quantum optimal control theory based on the Krotov method
to implement single-qubit $X$ and $Z$ gates and
two-qubit CNOT gates for inductively coupled
superconducting flux qubits with fixed qubit transition frequencies
and fixed off-diagonal qubit-qubit coupling. 
Our scheme that shares the same advantage of other directly
coupling schemes requires no additional coupler subcircuit and control
lines. The control lines needed are only for the manipulation of
individual qubits (e.g., a time-dependent magnetic flux or field
applied on each qubit). The
qubits are operated at the optimal coherence points 
and the gate operation times (single-qubit gates $< 1$ ns; CNOT 
gates $\sim 2$ ns) are
much shorter than the corresponding qubit decoherence time. 
A CNOT gate or other general quantum gates can be
implemented in a single run of pulse sequence rather than being
decomposed into several single-qubit and some entangled two-qubit
operations in series by composite pulse sequences.
Quantum gates constructed via our scheme are all
with very high fidelity 
(very low error) as our optimal control scheme takes into account the
fixed qubit detuning and fixed two-qubit
interaction as well as all other time-dependent magnetic-field-induced
single-qubit interactions and two-qubit couplings. 
The effect of leakage to higher energy-level states and the effect of
qubit decoherence on the quantum gate operations are also discussed. 

\end{abstract}

\maketitle

\section{Introduction}
Superconducting Josephson junction devices and circuits are proving to
be promising systems for quantum information processing \cite{You2005P,Clarke2008}. Due to the
great controllability of the qubits and microwaves in the
superconducting circuit
systems, single-qubit
\cite{Nakamura1999,Vion2002,Chiorescu2003,Bertet2005,Wallraff2005,Martinis2005,Yoshihara2006,Kakuyanagi2007,Chow2010,Paik2011,Corcoles2011,Rigetti2012},
two-qubit
\cite{Yamamoto2003,Sillanpaa2011,Majer2011,Niskanen07,Plantenberg07,DiCarlo2009,Chow2011,Dewes2012,Chow2012},
and three-qubit \cite{DiCarlo2010} operations have been
experimentally demonstrated.
The next natural step is to develop robust,
high-fidelity and scalable gates for larger scale quantum computation.

An essential prerequisite for quantum information processing and
quantum computation is precise coherent control of quantum systems or
quantum bits (qubits). Here, we focus our discussion on high-fidelity
quantum control for superconducting flux qubit systems.
The flux qubits have an advantage over other types of superconducting
qubits in the larger anharmonic energy level structure,
i.e. the difference between adjacent transition frequencies is larger
\cite{Lu12} and thus less leakage to higher energy-level states. 
Several schemes to implement local qubit operations and controllable
couplings using microwaves for multi-flux-qubit
systems have been proposed or/and realized. 
These schemes may be categorized into two groups: (i) directly coupling between
qubits \cite{Rigetti2005,Bertet2006,Liu06,Ashhab2006,Rigetti2010}
and (ii) indirectly coupling through an intermediate coupler
\cite{Niskanen07,Bertet2006,Niskanen2006,Harrabi2009,Grajcar2006}.
For example, a scheme to control the effective qubit-qubit interaction
for directly couping flux qubits with
fixed qubit transition frequencies and large detuning between the neighboring 
qubits using time-dependent magnetic fluxes (or microwaves) was proposed in
Ref.~\onlinecite{Liu06}.  While
this approach is advantageous due to the resonant nature of the
coupling, its disadvantage is that at least one of the qubits must be
biased away from the coherence optimal point \cite{Vion2002,Bertet2005,Yoshihara2006,Kakuyanagi2007}; this makes the qubit
susceptible to low-frequency flux noise and results in a shorter
coherence time.  
For the case of indirect coupling through an intermediate coupler, 
the coupling subcircuit or coupling nonlinear element can also be driven with
microwaves
\cite{Niskanen07,Bertet2006,Niskanen2006,Harrabi2009,Grajcar2006}. 
These indirectly coupling schemes have the advantage of enabling
the qubits to be operated at their optimal coherence points \cite{Vion2002,Bertet2005,Yoshihara2006,Kakuyanagi2007}. However,
in addition to the added circuit complexity, the effective couplings
between two qubits in the indirectly coupling schemes are generally
smaller so the two-qubit gating time is about
$10-200$ ns, at least one or two order(s) of magnitude longer than that of
directly coupling schemes.  
Schemes for microwave controllable coupling, which allow qubits to be
operated at optimal bias points and without intermediate
coupler subcircuit,
exist \cite{Rigetti2005,Bertet2006,Liu06,Ashhab2006,Rigetti2010}. But the effective qubit-qubit
couplings in these schemes are usually even smaller than those of the indirectly
coupling schemes. Smaller two-qubit coupling or equivalently longer
two-qubit gating time will usually make the qubits suffer more
decoherence effect during the gate operation, thus deteriorating the
gate fidelity.

The fidelities of single-qubit and two-qubit quantum gate operations
in a multi-qubit register for the present
directly/indirectly coupling schemes via microwaves in the unitary
case (without considering decoherence effect)
even under the two-level qubit
approximation are not perfect \cite{Niskanen07,Rigetti2005,Bertet2006,Liu06,Ashhab2006,Rigetti2010,Niskanen2006,Harrabi2009,Grajcar2006}. 
The microwave pulses 
for single-qubit gate operations
are generally obtained and applied under the approximation that the
qubit with large detuning with its neighboring qubits is effectively
decoupled with its neighbors even though there exist fixed
two-qubit couplings between them. 
But this
decoupling is only valid in the first order of a small parameter that is the
ratio of the two-qubit coupling to the detuning. 
For two-qubit operations, the microwave pulses
are commonly constructed and performed under the rotating wave
approximation, the adiabatic approximation for the nonlinear coupler,
or/and the 
approximation of neglecting other small residual two-qubit
interactions.
The qubit-decoupled approximation becomes excellent when the
effective qubit-qubit coupling is very small; however, a large qubit-qubit interaction
is favorable for two-qubit operations. 
Therefore, reaching a delicate balance between these two situations
results in, for instance, single-qubit gate errors greater than $10^{-3}$ for practical
experimental parameters even in the ideal, unitary case of simple two-level
approximation for each flux qubit. 
%(without considering leakage to higher energy states). 

Quantum optimal control theory is a powerful tool that provides a
variational framework for finding 
optimal control field profiles or sequences
by maximizing a desired physical objective (or minimizing a physical cost
function) within certain constraints
\cite{Rabitz88,Tannor92,Kosloff02,Khaneja05,Sporl07,Tannor11,Montangero2007,Nielsen08,potz2008,potz2009,Jirari2009,Rebentrost2009,Hwang2012,Tai2014}. 
%\cite{Rabitz88,Tannor92,Kosloff02,Khaneja05,Tannor11,Montangero2007,Nielsen08,Hwang2012}. 
In this paper, we apply the quantum optimal control theory to find
the control pulse sequences of externally applied ac magnetic fluxes
(fields) to implement fast and high-fidelity one-qubit and two-qubit gates on
superconducting flux qubits. 
There have been optimal control studies of quantum gates for
superconducting qubits, but all focusing on the gate operations of
Cooper-pair-box charge qubit systems \cite{Sporl07,Montangero2007,potz2008,potz2009,Jirari2009}.
Our investigation here is, to our knowledge, the first optimal control
study for the flux qubit systems. 
The flux qubits we consider have fixed
direct qubit-qubit couplings and fixed
transition frequencies but with large detuning
between neighboring qubits, 
which ensures the qubits are effectively
decoupled (to the first order) in the absence of time-dependent
control fields or signals.  
Our optimal control scheme requires no additional bias or control
lines beyond those used for the manipulation of individual qubits.   
In the absence of the time-dependent
magnetic fluxes, the flux qubits in our scheme 
are biased at the coherence optimal
points \cite{Vion2002,Bertet2005,Yoshihara2006,Kakuyanagi2007} to reduce dephasing due to flux noise, i.e.,to have a longer
coherence time. 
Furthermore, our optimal control scheme takes into account two-qubit
interaction, qubit detuning
and other time-dependent  magnetic field induced residual single-qubit
and two-qubit interactions when 
performing single-qubit and two-qubit gates. 
Thus the gate fidelities (errors) in the
unitary (closed-system) case can be as
high (low) as one wishes, limited only by the accuracy of the
approximated two-level qubit
Hamiltonian used and by the machine precision of the computation.
Besides,
the two-qubit CNOT gate operations of our scheme is about several folds to
 two orders of magnitude faster than the directly or/and indirectly
microwave controllable 
coupling schemes 
\cite{Niskanen07,Rigetti2005,Bertet2006,Liu06,Ashhab2006,Rigetti2010,Niskanen2006,Harrabi2009,Grajcar2006}.
% \cite{Rabitz88,Tannor92,Kosloff02,Khaneja05,Sporl07,Tannor11,Montangero2007,Nielsen08,potz2008,potz2009,Jirari2009,Rebentrost2009,Hwang2012}. 
By considering leakage to higher-energy-level states, the gate errors for
single-qubit gates using
the optimal control pulse sequences obtained are in the order of $10^{-8}$ and are in the order
of $10^{-6}$ for two-qubit CNOT gates.
To take the effect of qubit decoherence into account, we model the
qubit dynamics by a quantum master
equation with experimentally available relaxation and dephasing rates. 
The gate errors 
%by substituting the optimal pulse sequences
%obtained from the unitary case into 
in the presence of decoheence by considering the
master equation are still in the order of $10^{-6}$ for
single-qubit gates and in the order
of $10^{-5}$ for two-qubit CNOT gates. These gate errors are still
below the error threshold $10^{-4}$ 
($10^{-3}$ in \cite{Aliferis2009}; $10^{-2}$ if surface code error
correction is used \cite{Wang11,Fowler12L,Fowler12}) 
required for fault-tolerant quantum computation.

The paper is organized as follows. We first describe in
Sec.~\ref{sec:Hamiltonian} the
Hamiltonian of two inductively 
coupled flux qubits with each individual qubit controlled by a
time-dependent magnetic flux (field). 
The reduced Hamiltonian at the optimal bias point expressed in terms of
the two-level qubit basis states is then obtained.  
In Sec.~\ref{sec:QOCT}, a brief description about the
 quantum optimal control theory for performing state-independent
 quantum gate operations is presented.  
The control field pulse sequences obtained by the optimal control
theory and their corresponding gate errors and state evolutions 
are presented in Sec.~\ref{sec:Ressults}. It is also shown that 
our optimal control scheme is notably robust against leakage to states
outside the computational basis state space. 
The effect of qubit decoherence on gate errors is also discussed. 
A short conclusion with discussions of how to implement the optimal
control pulses experimentally is given in Sec.~\ref{sec:Conclusion}.

\section{Hamiltonian of Coupled Flux Qubits}
\label{sec:Hamiltonian}

\begin{figure}
\centering

\includegraphics[angle=270,width=0.5\textwidth]{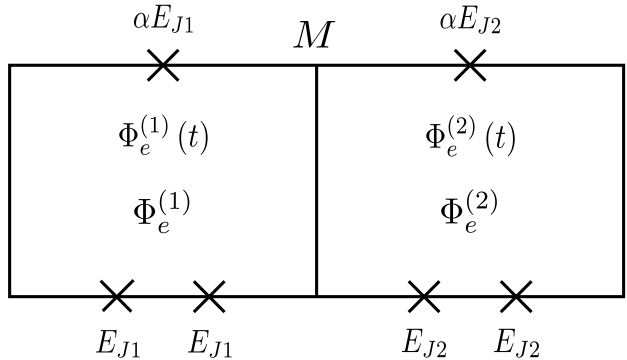}

\caption{\label{fig:coupled_flux_qubits} (Color online) Schematic
  illustration of two 
inductively coupled superconducting flux qubits.  The mutual
inductance is $M$ and there are three
junctions in each qubit loop.
The external static and time-dependent magnetic fluxes through the $l$th qubit
are denoted as $\Phi_e^{(l)}$ and $\Phi_e^{(l)}(t)$, respectively.
}
\end{figure}

 The system we consider here is two flux qubits right next 
to each other
%  sharing one leg of their respective loops 
%and coupled to each other  
and coupled by their mutual inductance $M$
  \cite{Majer2005,Liu06} as schematically 
illustrated in Fig. \ref{fig:coupled_flux_qubits}.
%which is due to energy stored in the magnetic field and the kinetic
%energy of the Cooper pairs \cite{Majer2005}.}
Each qubit loop contains three Josephson junctions \cite{Orlando1999,Chiorescu2003}, and one of them has
an area $\alpha$ times smaller than that of the two identical
junctions.
The larger Josephson junctions in the $l$th qubit loop have the Josephson
energy $E_{J,l}$ and the corresponding critical current
$I_{0}^{\left(l\right)}\equiv2\pi E_{J,l}/\Phi_{0}$, where
$\Phi_{0}=h/2e$ is the flux quantum
with the Planck constant $h$ and the elementary charge $e$. 
The capacitances
in the $l$th qubit loop satisfy the conditions $C_{1}^{\left(l\right)}=C_{2}^{\left(l\right)}=C_{J,l}$
and $C_{3}^{\left(l\right)}=\alpha C_{J,l}$, where
$C_{1}^{\left(l\right)}=C_{2}^{\left(l\right)}$ are the capacitance of
the larger Josephson junctions. The coupled flux qubit system considered here is
exactly the same as that discussed in Ref.~\cite{Liu06}.
Fluxoid quantization around each qubit loop imposes a constraint on
the phase drop across the three junctions \cite{Liu06}:  
% with the phase constraint condition 
\begin{equation}
\sum_{i}\varphi_{i}^{\left(l\right)}+2\pi\frac{\Phi_{e}^{\left(l\right)}}{\Phi_{0}}+2\pi\frac{\Phi_{e}^{\left(l\right)}\left(t\right)}{\Phi_{0}}=0,
\label{phase}
\end{equation}
where $\varphi_{i}^{\left(l\right)}$ is the gauge-invariant phase of each
junction,
$\Phi_{e}^{\left(l\right)}$ and $\Phi_{e}^{\left(l\right)}\left(t\right)$
are, respectively, the static (dc) and  
 time-dependent magnetic fluxes applied through the $l$th qubit.
Here the self-inductance is
considered negligible as compared to the Josephson inductance. 
The requirement for our coupled-qubit system is to have
  suitable mutual inductance for qubits placed next to each other as
illustrated in Fig. \ref{fig:coupled_flux_qubits}
and independent flux lines
  to enable control over each individual qubit with external dc bias
  magnetic fluxes and time-dependent control magnetic fluxes. 
The qubit-loop size for three-junction
  flux qubits could range experimentally
  from submicrometer to micrometer.
As long as the design of the qubit loops fulfills the above requirement, the
  actual size of each qubit loop or the distance between the qubits
  can have some flexibility.

The total Hamiltonian of the two coupled three-junction flux qubits
reads \cite{Orlando1999,Liu06}:
\begin{eqnarray}
H & = & MI_{1}I_{2}+\sum_{l}\left\{ \frac{P_{P,l}^{2}}{2M_{P,l}}+\frac{P_{Q,l}^{2}}{2M_{Q,l}}\right.\nonumber \\
 &  & \qquad \qquad \quad +2\pi\dot{f}_{c}^{\left(l\right)}\left(t\right)\left(\frac{\alpha}{1+2\alpha}\right)P_{P,l}+2E_{J,l}\left(1-\cos\varphi_{Q}^{\left(l\right)}\cos\varphi_{P}^{\left(l\right)}\right)\nonumber \\
 &  & \qquad \qquad \quad \left.+\alpha
   E_{J,l}\left[1-\cos\left(2\varphi_{P}^{\left(l\right)}+2\pi
       f_{l}+2\pi
       f_{c}^{\left(l\right)}\left(t\right)\right)\right]\right\}, 
\label{eq:Htot}
\end{eqnarray}
with the redefined phases $\varphi_{P}^{\left(l\right)}=(\varphi_{1}^{\left(l\right)}+\varphi_{2}^{\left(l\right)})/2$,
$\varphi_{Q}^{\left(l\right)}=(\varphi_{1}^{\left(l\right)}-\varphi_{2}^{\left(l\right)})/2$,
the reduced dc bias magnetic flux $f_{l}=\Phi_{e}^{\left(l\right)}/\Phi_{0}$,
and the reduced time-dependent control magnetic flux $f_{c}^{\left(l\right)}\left(t\right)=\Phi_{e}^{\left(l\right)}\left(t\right)/\Phi_{0}$.
The effective masses are $M_{Q,l}=2\left(\Phi_{0}/2\pi\right)^{2}C_{J,l}$
and $M_{P,l}=\left(1+2\alpha\right)M_{Q,l}$, which correspond to
the effective momenta $P_{Q,l}=-\imath\hbar(\partial/\partial\varphi_{Q}^{\left(l\right)})$
and
$P_{P,l}=-\imath\hbar(\partial/\partial\varphi_{P}^{\left(l\right)})$.
%where $C_{J,l}$ is the capacitance of
%the larger Josephson junctions in the $l$th qubit loop and $\alpha$ is 
%the area ratio of the smaller junction to the larger junctions in a 
%qubit loop,
The persistence current in the $l$th qubit loop is
\begin{eqnarray}
I_{l} & = & \frac{\alpha I_{0}^{\left(l\right)}}{1+2\alpha}\left[\sin\left(\varphi_{P}^{\left(l\right)}+\varphi_{Q}^{\left(l\right)}\right)+\sin\left(\varphi_{P}^{\left(l\right)}-\varphi_{Q}^{\left(l\right)}\right)\right.\nonumber \\
 &  &\qquad \quad \quad \left.+\sin\left(2\varphi_{P}^{\left(l\right)}+2\pi f_{l}+2\pi f_{c}^{\left(l\right)}\left(t\right)\right)\right].
\label{eq:I}
\end{eqnarray}
Here the self-interaction
terms of the external time-dependent control magnetic fields 
$\sum_{l=1}^{2}\left(\alpha/2\right)C_{J,l}[\dot{\Phi}_{e}^{\left(l\right)}(t)]^{2}$
is neglected since they vary only the global phase of the state
vector and do not affect the control of the coupled flux qubits. 
To simulate the dynamics of the coupled flux-qubit system with
Hamiltonian Eq.~(\ref{eq:Htot}), 
the parameter values of the mutual conductance $M$ as well as 
the Josephson energies
of the junctions $E_{J,l}$,
 the capacitance of
the larger Josephson junctions $C_{J,l}$ and 
the area ratio $\alpha$ of the smaller junction to one of the larger junctions in each 
qubit loop are required. 
These parameter values
can be obtained through the experimental measurement and
characterization of the system.   
The actual values of these parameters we choose for our simulations are given in
Sec.~\ref{sec:Ressults}. They are all experimentally available or
realistic values.

We will keep the time-dependent control amplitudes small such that the
reduced time-dependent magnetic flux satisfies
$|f_{c}^{(l)}(t)|=|\Phi_{e}^{(l)}(t)/\Phi_{0}|\lesssim 10^{-3}$. 
Small time-dependent control amplitude allows the approximation of  
$\sin[2\pi f_{c}^{\left(l\right)}(t)]\sim2\pi f_{c}^{\left(l\right)}(t)$
and $\cos[2\pi f_{c}^{\left(l\right)}(t)]\sim 1$
when expanding the last term of the Hamiltonian of
Eq.~(\ref{eq:Htot}). As a result, the time-dependent Hamiltonian becomes
linear in $f_c^{(l)}(t)$ or $f_c^{(1)}(t)f_c^{(2)}(t)$, which can be
readily incorporated with the Krotov quantum optimal control method
that we will employ later. 
This weak-amplitude approximation of the control fields also keeps the
qubit not deviating much 
from the dc bias point that is set to be the optimal coherence
point in our case. Moreover, this weak-amplitude approximation  
helps reduce unwanted possible 
excitations to the higher-energy-level states outside the
computational state space 
when we make two-level (qubit) approximation discussed later.

Making this weak-amplitude approximation, we obtain
from Eq.~(\ref{eq:Htot}) the Hamiltonian
similar to that in Ref.~\cite{Liu06} as
\begin{equation}
H=\sum_{l=1}^{2}\left(H_{l}+H_{D}^{\left(l\right)}\right)+\sum_{l\neq m=1}^{2}H_{lm}+H_{C}+H_{A}.\label{eq:Hamiltonian02}
\end{equation}
The first term $H_{l}$ is the single-qubit Hamiltonian and reads:
\begin{eqnarray}
H_{l} & = & \frac{P_{P,l}^{2}}{2M_{P,l}}+\frac{P_{Q,l}^{2}}{2M_{Q,l}}+2E_{J,l}\left(1-\cos\varphi_{Q}^{\left(l\right)}\cos\varphi_{P}^{\left(l\right)}\right)\nonumber \\
 &  & +\alpha E_{J,l}\left[1-\cos\left(2\varphi_{P}^{\left(l\right)}+2\pi f_{l}\right)\right].\label{eq:Hamiltonian03}
\end{eqnarray}
The Hamiltonian $H_{D}^{\left(l\right)}$ plays the role of a driving Hamiltonian
representing the interaction between the $l$th qubit and its time-dependent magnetic field. It takes the form
\begin{equation}
H_{D}^{\left(l\right)}=2\pi\alpha\left[f_{c}^{\left(l\right)}\left(t\right)E_{J,l}\sin\left(2\varphi_{P}^{\left(l\right)}+2\pi
    f_{l}\right)+\dot{f}_{c}^{\left(l\right)}\left(t\right)\frac{P_{P,l}}{1+2\alpha}\right].
\label{driving_Ham}
\end{equation}
The last three terms in Eq. (\ref{eq:Hamiltonian02}) come from the
inductive coupling between the two flux qubits. The Hamiltonian
$H_{lm}$ describes the qubit-qubit 
interaction controlled
by one of the time-dependent magnetic flux 
%[$\Phi_{e}^{\left(1\right)}\left(t\right)$ or
%$\Phi_{e}^{\left(2\right)}\left(t\right)$; 
[$f_{c}^{\left(1\right)}\left(t\right)$ or
$f_{c}^{\left(2\right)}\left(t\right)$]  
and is written as
\begin{equation}
H_{lm}=-2\pi M\left(\frac{\alpha}{1+2\alpha}\right)I^{\left(l\right)}I_{0}^{\left(m\right)}f_{c}^{\left(m\right)}\left(t\right)\cos\left(2\varphi_{P}^{\left(m\right)}+2\pi f_{m}\right),
\end{equation}
where 
\begin{eqnarray}
I^{\left(l\right)} & = & \frac{\alpha I_{0}^{\left(l\right)}}{1+2\alpha}\left[\sin\left(\varphi_{P}^{\left(l\right)}+\varphi_{Q}^{\left(l\right)}\right)+\sin\left(\varphi_{P}^{\left(l\right)}-\varphi_{Q}^{\left(l\right)}\right)\right.\nonumber \\
 &  & \qquad \quad \quad \left.-\sin\left(2\varphi_{P}^{\left(l\right)}+2\pi
     f_{l}\right)\right]
\label{loop_current}
\end{eqnarray}
is the loop current of the $l$th qubit when no
time-dependent magnetic fluxes are applied [cf.,
Eq.~(\ref{eq:I})]. The qubit-qubit interaction $H_{C}$ 
controlled by two 
simultaneously applied time-dependent magnetic fluxes through,
respectively, the two qubits is  
\begin{equation}
H_{C}=M\left(\frac{2\pi\alpha}{1+2\alpha}\right)^{2}\prod_{l=1}^{2}I_{0}^{\left(l\right)}f_{c}^{\left(l\right)}\left(t\right)\cos\left(2\varphi_{P}^{\left(l\right)}+2\pi f_{l}\right).
\end{equation}
We still keep this term, although it is much smaller than the term
controlled by a single time-dependent magnetic flux. 
The Hamiltonian 
\begin{equation}
H_{A}=MI^{\left(1\right)}I^{\left(2\right)},
\end{equation}
describes an always-on interaction between the two flux qubits without
time-dependent magnetic fluxes being applied, where
$I^{\left(l\right)}$ is defined in Eq.~(\ref{loop_current}).

The Hamiltonian, Eq.~(\ref{eq:Hamiltonian02}), in the two-qubit
computational state basis $\left\{ \left|g_{1}\right\rangle ,\left|e_{1}\right\rangle \right\} \otimes\left\{ \left|g_{2}\right\rangle ,\left|e_{2}\right\rangle \right\} $,
where $\left|g_{l}\right\rangle $ and $\left|e_{l}\right\rangle $
are the lowest two energy-level states of $H_{l}$
of Eq.~(\ref{eq:Hamiltonian03}), becomes
\begin{eqnarray}
\frac{H}{\hbar} & = &
\sum_{l=1}^{2}\left[\frac{\omega_{l}}{2}\sigma_{z}^{\left(l\right)}+\kappa_{1}^{\left(l\right)}(t)\sigma_{z}^{\left(l\right)}+\kappa_{2}^{\left(l\right)}(t)\sigma_{x}^{\left(l\right)}\right]
\nonumber \\
 &  &-\sum_{l\neq m=1}^{2}\left[\chi_{1}^{\left(lm\right)}(t)\sigma_{z}^{\left(l\right)}+\chi_{2}^{\left(lm\right)}(t)\sigma_{x}^{\left(l\right)}\right]\nonumber \\
 &  & -\sum_{l\neq
   m=1}^{2}\left[\Xi_{11}^{\left(lm\right)}(t)\sigma_{z}^{\left(l\right)}\sigma_{z}^{\left(m\right)}+\Xi_{22}^{\left(lm\right)}(t)\sigma_{x}^{\left(l\right)}\sigma_{x}^{\left(m\right)}\right. \nonumber
 \\
 &  &\qquad \qquad \left. +\Xi_{12}^{\left(lm\right)}(t)\sigma_{z}^{\left(l\right)}\sigma_{x}^{\left(m\right)}+\Xi_{21}^{\left(lm\right)}(t)\sigma_{x}^{\left(l\right)}\sigma_{z}^{\left(m\right)}\right]\nonumber \\
 &  & +\Theta_{11}(t)\sigma_{z}^{\left(1\right)}\sigma_{z}^{\left(2\right)}+\Theta_{22}(t)\sigma_{x}^{\left(1\right)}\sigma_{x}^{\left(2\right)}+\Theta_{12}(t)\sigma_{z}^{\left(1\right)}\sigma_{x}^{\left(2\right)}+\Theta_{21}(t)\sigma_{x}^{\left(1\right)}\sigma_{z}^{\left(2\right)}\nonumber \\
 &  & +\Lambda_{11}\sigma_{z}^{\left(1\right)}\sigma_{z}^{\left(2\right)}+\Lambda_{22}\sigma_{x}^{\left(1\right)}\sigma_{x}^{\left(2\right)}+\Lambda_{12}\sigma_{z}^{\left(1\right)}\sigma_{x}^{\left(2\right)}+\Lambda_{21}\sigma_{x}^{\left(1\right)}\sigma_{z}^{\left(2\right)},\label{eq:Hamiltonian04}
\end{eqnarray}
where the single-qubit transition frequencies $\omega_{l}$ and the driving
Hamiltonian amplitudes $\kappa_{1}^{\left(l\right)}(t)$ and
$\kappa_{2}^{\left(l\right)}(t)$ are 
\begin{equation}
\omega_{l}=\frac{1}{\hbar}\left(\left\langle
    e_{l}\right|H_{l}\left|e_{l}\right\rangle -\left\langle
    g_{l}\right|H_{l}\left|g_{l}\right\rangle \right),
\label{transition_freq}
\end{equation}
\begin{eqnarray}
\kappa_{1}^{\left(l\right)}(t) & = & \frac{1}{2\hbar}\left(\left\langle
    e_{l}\right|H_{D}^{\left(l\right)}\left|e_{l}\right\rangle
  -\left\langle g\right|H_{D}^{\left(l\right)}\left|g_{l}\right\rangle
\right),
\label{kappa_1}
\end{eqnarray}
\begin{eqnarray}
\kappa_{2}^{\left(l\right)}(t)& = & \frac{1}{\hbar}\left\langle e_{l}\right|H_{D}^{\left(l\right)}\left|g_{l}\right\rangle ,
\end{eqnarray}
respectively, the other
time-dependent controllable interaction strengths, 
$\chi_{1}^{\left(lm\right)}(t)$, $\chi_{2}^{\left(lm\right)}(t)$,
$\Xi_{ij}^{\left(lm\right)}(t)$ and $\Theta_{ij}(t)$, coming from
qubit-qubit inductive interaction are
\begin{equation}
\chi_{1}^{\left(lm\right)}(t)=\frac{2\pi}{\hbar}\beta_{M}f_{c}^{\left(l\right)}\left(t\right)\Omega_{1}^{\left(l\right)}\Delta^{\left(m\right)},\label{eq:Param01}
\end{equation}
\begin{equation}
\chi_{2}^{\left(lm\right)}(t)=\frac{2\pi}{\hbar}\beta_{M}f_{c}^{\left(l\right)}\left(t\right)\Omega_{2}^{\left(l\right)}\Delta^{\left(m\right)},
\end{equation}
\begin{equation}
\Xi_{ij}^{\left(lm\right)}(t)=\frac{2\pi}{\hbar}\beta_{M}f_{c}^{\left(l\right)}\left(t\right)\Omega_{i}^{\left(l\right)}\lambda_{j}^{\left(m\right)},
\end{equation}
\begin{equation}
\Theta_{ij}(t)=\frac{\left(2\pi\right)^{2}}{\hbar}\beta_{M}f_{c}^{\left(1\right)}\left(t\right)f_{c}^{\left(2\right)}\left(t\right)\Omega_{i}^{\left(1\right)}\Omega_{j}^{\left(2\right)},
\end{equation}
respectively, and the static fixed  qubit-qubit interaction strengths
 $\Lambda_{ij}$ are
\begin{equation}
\Lambda_{ij}=\frac{\beta_{M}}{\hbar}\lambda_{i}^{\left(1\right)}\lambda_{j}^{\left(2\right)}.\label{eq:Param02}
\end{equation}
The relevant parameters in Eqs.~(\ref{eq:Param01})-(\ref{eq:Param02})
are $\beta_{M}= MI_{0}^{\left(1\right)}I_{0}^{\left(2\right)}$
which is the mutual inductive energy with respect to the 
%critical current $I_{0}^{\left(l\right)}$ 
critical current $I_{0}^{\left(l\right)}=2\pi E_{J,l}/\Phi_{0}$
of the larger Josephson junctions in
each qubit loop,
\begin{equation}
\lambda_{1}^{\left(l\right)}=\frac{1}{2}\left(\left\langle e_{l}\right|\frac{I^{\left(l\right)}}{I_{0}^{\left(l\right)}}\left|e_{l}\right\rangle -\left\langle g_{l}\right|\frac{I^{\left(l\right)}}{I_{0}^{\left(l\right)}}\left|g_{l}\right\rangle \right),
\label{lambda1}
\end{equation}
\begin{equation}
\lambda_{2}^{\left(l\right)}=\left\langle e_{l}\right|\frac{I^{\left(l\right)}}{I_{0}^{\left(l\right)}}\left|g_{l}\right\rangle ,
\label{lambda2}
\end{equation}
\begin{eqnarray}
\Omega_{1}^{\left(l\right)} & = & \frac{1}{2}\left[\left\langle e_{l}\right|\left(\Upsilon^{\left(l\right)}-\frac{I^{\left(l\right)}}{I_{0}^{\left(l\right)}}\right)\left|e_{l}\right\rangle -\left\langle e_{l}\right|\left(\Upsilon^{\left(l\right)}-\frac{I^{\left(l\right)}}{I_{0}^{\left(l\right)}}\right)\left|e_{l}\right\rangle \right],
\label{Omega1}
\end{eqnarray}
\begin{equation}
\Omega_{2}^{\left(l\right)}=\left\langle e_{l}\right|\left(\Upsilon^{\left(l\right)}-\frac{I^{\left(l\right)}}{I_{0}^{\left(l\right)}}\right)\left|g_{l}\right\rangle ,
\label{Omega2}
\end{equation}
and
\begin{eqnarray}
\Delta^{\left(l\right)} & = & \frac{1}{2}\left[\left\langle e_{l}\right|\Upsilon^{\left(l\right)}\left|e_{l}\right\rangle +\left\langle g_{l}\right|\Upsilon^{\left(l\right)}\left|g_{l}\right\rangle \right],
\label{Delta}
\end{eqnarray}
where the operator $\Upsilon^{\left(l\right)}$ is defined as
\begin{equation}
\Upsilon^{\left(l\right)}\equiv\frac{\alpha}{1+2\alpha}\cos\left(2\varphi_{P}^{\left(l\right)}+2\pi
  f_{l}\right)+\frac{I^{\left(l\right)}}{I_{0}^{\left(l\right)}}.
\label{upsilon}
\end{equation}
The Hamiltonian Eq.~(\ref{eq:Hamiltonian04})
with parameters defined in
Eqs.~(\ref{transition_freq})-(\ref{upsilon}) is valid
for weak time-dependent magnetic fluxes
(fields) and arbitrary static 
bias magnetic fluxes. If we take the time dependence of the control
magnetic flux 
to be sinusoidal, i.e., $e^{\pm i\omega t}$, and making the relevant
rotating wave approximation, we can revert to the Hamiltonian of 
Ref.~\cite{Liu06}. 
Different from Ref.~\cite{Liu06}, our control scheme, however, does not require qubits being biased
away from the optimal coherence points.
% at which the reduced static or dc
%bias magnetic fluxes are
%$f_{l}=\Phi_{e}^{\left(l\right)}/\Phi_{0}=0.5$. 
In contrast, we set the reduced static or dc
bias magnetic fluxes to be $f_{1}=f_{2}=0.5$ so that the qubits 
are at the optimal coherence points and are thus insensitive
to low-frequency flux noise in the first order. 
At the optimal coherence points of $f_{1}=f_{2}=0.5$, Hamiltonian  $H_{l}$
of Eq.~(\ref{eq:Hamiltonian03}) is invariant under the
parity transformation of $\varphi_{P}^{\left(l\right)}$, i.e.,
$H_{l}(-\varphi_{P}^{\left(l\right)})=H_{l}(\varphi_{P}^{\left(l\right)})$. Thus 
its eigenstates have definite parities and the lowest two energy-level
eigenstates have the opposite parities \cite{Liu2005,Deppe2008}. 
Since the Hamiltonian of Eq.~(\ref{driving_Ham}) and the loop currents of
Eq.~(\ref{loop_current}) are odd functions
of $\varphi_{P}^{\left(l\right)}$ and the operators
$\Upsilon^{\left(l\right)}-(I^{\left(l\right)}/I_{0}^{\left(l\right)})$
from Eq.~(\ref{upsilon}) are even functions
of $\varphi_{P}^{\left(l\right)}$ at $f_{1}=f_{2}=0.5$, the parameters  $\kappa_{1}^{\left(l\right)}(t)$,
$\lambda_{1}^{\left(l\right)}$ and $\Omega_{2}^{\left(l\right)}$
defined in  Eqs.~(\ref{kappa_1}), (\ref{lambda1}) and  (\ref{Omega2}), respectively, vanish
due to the parity symmetry consideration. 
As a consequence, most parameters in Eqs.~(\ref{eq:Param01})-(\ref{eq:Param02})
vanish, and Eq (\ref{eq:Hamiltonian04}) at the optimal bias points
simplifies to
\begin{eqnarray}
\frac{H}{\hbar} & = & \sum_{l=1}^{2}\left[\frac{\omega_{l}}{2}\sigma_{z}^{\left(l\right)}+\kappa_{2}^{\left(l\right)}(t)\sigma_{x}^{\left(l\right)}\right]+\Lambda_{22}\sigma_{x}^{\left(1\right)}\sigma_{x}^{\left(2\right)}\nonumber \\
 &  & -\sum_{l\neq m=1}^{2}\left[\chi_{1}^{\left(lm\right)}(t)\sigma_{z}^{\left(l\right)}+\Xi_{12}^{\left(lm\right)}(t)\sigma_{z}^{\left(l\right)}\sigma_{x}^{\left(m\right)}\right]+\Theta_{11}(t)\sigma_{z}^{\left(1\right)}\sigma_{z}^{\left(2\right)}.\label{eq:Hamiltonian05}
\end{eqnarray}
We will use Eq.~(\ref{eq:Hamiltonian05}) combining with quantum
optimal control theory to find control magnetic field pulses
$f_c^{(l)}(t)$ for the implementation of high-fidelity one- and two-qubit gates.

\section{Krotov Quantum Optimal Control Method} 
\label{sec:QOCT}

Quantum optimal control theory enables us to
realize accurate state-independent quantum gates by selecting optimal pulse
shapes (arbitrarily shaped pulses and duration; or continuous
dynamical modulation) for the external control within experimental
capabilities
\cite{Khaneja05,Sporl07,Montangero2007,potz2008,potz2009,Jirari2009,Rebentrost2009,Hwang2012,Tai2014}.
%For the quantum gate operations investigated here, we need 
To perform state-independent optimal control, the equation of motion
for the time evolution operator (or propagator) $U(t)$ is needed: 
\begin{equation}
\imath\hbar\frac{\partial}{\partial
  t}U\left(t\right)
={H}\left[t,\epsilon\left(t\right)\right]U\left(t\right),
\label{eq:Sch1} 
\end{equation}
where $H$ is the system Hamiltonian and $\epsilon(t)$ is a
time-dependent control field.
We choose the trace distance
between the desired target gate operation $O$
and the actual (could be nonunitary) propagator $U(T)$
at the final operation time $T$
to characterize the gate error:
\begin{eqnarray}
\eta&=&\frac{1}{2N}
  \text{Tr}\left\{\left[O-U\left(T\right)\right]^{\dagger}\left[O-U(T)\right]\right\} 
\label{error1},
%\\
%&=&\frac{1}{2N}
%\left\langle \left\langle
%    U\left(T\right)-O\left|U\left(T\right)-O\right\rangle
%  \right\rangle \right..
%\label{error2}
\end{eqnarray}
where $N$ is the dimension of the matrix $U(t)$. If the actual
time evolution operator $U(T)$ is equal to the target gate operation
$O$, then $\eta=1-\mathcal{F}=0$ 
with $\mathcal{F}$ denoting the gate fidelity.
Sometimes, one is also interested in calculating the errors of the
operations (operators or
matrices) in a subspace of a unitary matrix. 
When projection is made to the subspace, the projected
matrix in the subspace is in general no longer unitary.  
For instance, when the evolution operator of the multi-level coupled
flux-qubit system  $U(T)$ for the
multi-level Hamiltonian,  Eq.~(\ref{eq:Htot}), is projected into the 
approximated two-level qubit computational state subspace, the
projected time evolution operator is no longer unitary.  
In this case, Eq.~(\ref{error1}) with slight modification 
[see Eq.~(\ref{errorP})] is still an appropriate measure for
calculating the gate error.

%In this case, the error of $(1-\mathcal{F})$, with $\mathcal{F}$ defined
%in Eq.~(\ref{fidelity1}), 
%may not provide all the information on quantifying the difference
%between two operators (matrices), one of which is not unitary.
%Hence, we choose 
%the trace distance
%between the desired target gate operation $O$
%and the actual (nonunitary) propagator $U(T)$
%at the final operation time $T$
%to characterize the gate error:
%\begin{eqnarray}
%\eta&=&\frac{1}{2N}
%  \text{Tr}\left\{\left[O-U\left(T\right)\right]^{\dagger}\left[O-U(T)\right]\right\} 
%\label{error1}.
%\\
%&=&\frac{1}{2N}
%\left\langle \left\langle
%    U\left(T\right)-O\left|U\left(T\right)-O\right\rangle
%  \right\rangle \right..
%\label{error2}
%\end{eqnarray}
%One can show that the gate error $\eta$ defined in
%Eq.~(\ref{error1}) reduces to $(1-\mathcal{F})$ with $\mathcal{F}$ defined
%in Eq.~(\ref{fidelity1}) when $U(T)$ is unitary.  
%The time evolution operator here is the one projecting on
%the computational space, and 

In realistic control problems, 
it is desirable that the optimal control sequence can provide highest
quality (fidelity) with minimum energy consumption. 
Therefore, we define the cost function for our optimal control problem
as
\begin{equation}
J=\eta
+\int_{0}^{T}\frac{\lambda}{S\left(t\right)}\left[\epsilon\left(t\right)-\epsilon_{\text{ref}}\left(t\right)\right]^{2}dt,
\label{eq:CostFun02}
\end{equation}
where $\eta$ is defined in Eq.~(\ref{error1}), 
 $S\left(t\right)$, a positive shape function, and
$\lambda$, a weight, can be adjusted and chosen
empirically \cite{Tannor92,Kosloff02}.
%and $\epsilon_{\rm ref}(t)$, a reference control value, can be properly
%chosen \cite{Tannor92,Kosloff02}. 
Here, the reference
field $\epsilon_{\text{ref}}\left(t\right)$ is chosen to be 
the control sequence in the previous iteration, 
i.e., $\epsilon_{\text{ref}}\left(t\right)=\epsilon^{\left(i\right)}\left(t\right)$ 
%in Eq.~(\ref{update_rule}), 
such that the control field energy constraint
in the cost function $J$ of Eq.~(\ref{eq:CostFun02}) has the physical
interpretation that the change 
of the control pulse energy in each iteration is limited
\cite{Tannor92,Kosloff02}. When the 
iterative procedure approaches the optimal solution, the
change in the control field is minimal or vanishing. 
Therefore, this choice of the reference
field $\epsilon_{\text{ref}}\left(t\right)$ ensures that the iterative
method is found to reduce the total objective $J$  of
Eq.~(\ref{eq:CostFun02}) by reducing 
the gate error $\eta$ rather than the total control pulse energy. 
With Eqs.~(\ref{eq:Sch1}) and (\ref{eq:CostFun02}) and the initial
propagator $U(0)=I$, where $I$ is the 
%$|U(0)\rangle\rangle=|I\rangle\rangle$, 
%where $|I\rangle\rangle$ is the 
identity operator, 
%in the column vector form, 
one can then investigate the
state-independent quantum gate optimal control problem.

We will use the Krotov iterative method for the quantum gate optimal control
\cite{Tannor92,Tannor11,Nielsen08,Hwang2012,Tai2014,krotov1996}. 
%adapted for quantum control by Tannor et al. \cite{Tannor92,Tannor93}.
The Krotov method has several
appealing advantages \cite{Tannor92,Tannor11,Nielsen08}
over the standard gradient optimization methods: 
(a) monotonic increase of the objective 
with iteration number,
(b) no requirement for a line search,  and (c) macrosteps
at each iteration.
The optimal algorithm following the Krotov method \cite{krotov1996} 
can be found in Refs.~\onlinecite{Tannor92,Kosloff02,Nielsen08,Hwang2012}.

\section{Quantum gate operations via quantum optimal control theory}
\label{sec:Ressults}

We use the reduced Hamiltonian, Eq.~(\ref{eq:Hamiltonian05}),
with two lowest energy-level states for each qubit to
obtain control sequences for
single-qubit $X$ and $Z$ gate operations and two-qubit CNOT
gate operations  by optimizing the cost function $J$,
Eq.~(\ref{eq:CostFun02}), using the Krotov iterative method. 
Simulations 
%with the obtained optimal control sequences 
on the two-qubit Hamiltonian, Eq.~(\ref{eq:Htot}), without
making the weak-amplitude approximation and with the lowest five
energy states for each qubit, 
using the optimal control field sequences found via
Eq.~(\ref{eq:Hamiltonian05}) will also be performed for
error comparison.
This allows us to test how well the optimal control pulses obtained
by the weak-amplitude approximation perform, and to determine how severe
the leakage to the higher-energy-level states is.
The target gate operation $O$ is defined in the two-qubit computational
state basis. The evolution operator or propagator $U(T)$ results from Eq.~(\ref{eq:Htot}) using 
the optimal control pulse sequences
$f_{c}^{\left(1\right)}\left(t\right)=\Phi_{e}^{\left(1\right)}\left(t\right)/\Phi_{0}$
and
$f_{c}^{\left(2\right)}\left(t\right)=\Phi_{e}^{\left(2\right)}\left(t\right)/\Phi_{0}$
obtained from Eq.~(\ref{eq:Hamiltonian05}) is in a larger multi-level-state
space. We thus define a projection operator $P$ to project $U(T)$ onto
the subspace subtended by the computational basis states, and compare
the target operator $O$ with the resultant propagator $PU(T)$ which is
in general no longer unitary. The
error in this case is defined from Eq.~(\ref{error1}) as 
\begin{equation}
\eta_P=(1/2N){\rm Tr}\left\{[O-PU(t)]^{\dagger}[O-PU(t)]\right\},
%\eta_P=(1/32){\rm Tr}\left\{[O-PU(t)]^{\dagger}[O-PU(t)]\right\}.
%\left\langle \left\langle
%    PU\left(T\right)-O\left|\right.PU\left(T\right)-O\right\rangle
%  \right\rangle.
\label{errorP}
\end{equation}
where $N$ is the dimension of the $O$ matrix or $PU(t)$ matrix.

%{\color{red} In order to simulate the coupled qubits system, we need the experimental values. The mutual inductance, $M$, the area ratio of the smaller junction, $\alpha$, and the Josephson energies of the junctions in each loop, $E_{J,l}$, should be known to obtain the coupling strength between the two qubits which is corresponding to the first term in Eq. (\ref{eq:Htot}). For the single-qubit Hamiltonian, the effective masses are obtained from the capacitance of the large junctions, $C_{J,l}$, and then the kinetic part in Eq. (\ref{eq:Hamiltonian03}) can be simulated numerically. Combining the kinetic part with the potential part which can be solved by the Josephson energies of the junctions, we can learn the single-qubit Hamiltonian. The effect of the driving terms can be simply tuned by the time-dependent magnetic fluxes which appear in the last term of Eq. (\ref{eq:Htot}), and the total Hamiltonian is also obtained after adding the coupling Hamiltonian. In order to transform the system into the computational basis, the transformation matrices are obtained by diagonalizing the single-qubit Hamiltonian.}
We choose the parameters in the two-qubit Hamiltonian with
experimentally available or realistic values \cite{Chiorescu2003}.
The Josephson energy of the larger junctions in the qubit loops is chosen to be  $E_{J,1}/\hbar=2E_{J,2}/\left(5\hbar\right)=E_{J}/\hbar=2\pi\times248.72$
GHz and the charging energy $E_{C,l}=e^{2}/2C_{J,l}$ is chosen to be $E_{J,l}/E_{C,l}=35$. The capacitance of the smaller
Josephson junction is $C_{3,l}=\alpha C_{J,l}$ with the ratio $\alpha=0.8$.
%and the mutual inductance is $M=1$ pH. 
To fit the value of the coupling strength $\Lambda_{22}=2\pi\times0.4\text{
  GHz}$ reported in the experiment of Ref.~\onlinecite{Izmalkov2004}, 
we take the mutual inductance $M=1$ pH and obtain
$\beta_{M}=3.75\times10^{-3}E_{J}$.
Then other parameters are determined by the calculations through the
Hamiltonian, Eq.~(\ref{eq:Htot}). We then obtain the qubit transition
energies
 $\omega_{1}=2\pi\times3.30\text{ GHz}$ and $\omega_{2}=2\pi\times8.24\text{ GHz}$,
and the single-qubit driving amplitudes $\kappa_{2}^{\left(1\right)}=-2\pi
f_{c}^{\left(1\right)}\left(t\right)\times1.02\times10^{3}\text{
  GHz}$ and
$\kappa_{2}^{\left(2\right)}=-2\pi
f_{c}^{\left(2\right)}\left(t\right)\times2.57\times10^{3}\text{
  GHz}$.
The time-dependent coefficients of the last three coupling terms in
Eq.~(\ref{eq:Hamiltonian05}) are also obtained to be  $\chi_{1}^{\left(lm\right)}=2\pi f_{c}^{\left(l\right)}\left(t\right)\times4.4\times10^{-3}\text{ GHz},$
$\Xi_{12}^{\left(lm\right)}=2\pi f_{c}^{\left(l\right)}\left(t\right)\times8.22\times10^{-2}\text{ GHz},$
and $\Theta_{11}=2\pi
f_{c}^{\left(1\right)}\left(t\right)f_{c}^{\left(2\right)}\left(t\right)\times1.66\times10^{-2}\text{
  GHz}$.
%these coefficients are very small when compared with 
%the qubit transition frequencies, qubit
%detuning, the single-qubit driving amplitudes and the fixed static 
%qubit-qubit interaction strength.
We note here that the dynamics and also the optimization results 
that will be presented later
depend on the values of the microscopic parameters.
These parameter values
can be obtained through the experimental measurement and
characterization of the system but always come with some error bars.    
Thus if the actual system is, for example, slightly away from the 
assumed control point (a set of values for the system parameters) chosen
for the quantum optimal control theory,  
then the optimal control pulse sequences sent to the experiment may 
result in gate fidelities not close to optimal. 
This issue of imprecise knowledge of the system parameters seems to
hinder the practical use and experimental applicability of the quantum
optimal control theory. 
Fortunately, there may be ways around, for example, using the 
adaptive model-free closed-loop feedback control method. We
postpone the
discussions regarding this until Sec.~\ref{sec:Conclusion}.

In the following, we first 
present our optimal control results for single-qubit and
two-qubit gates in the unitary case together with brief descriptions
of corresponding conventional approaches for comparison.  
The results considering the effect of leakage to
higher-energy-level states and the effect of qubit decoherence are
presented subsequently. 

\subsection{Single-qubit gate: unitary case}

\begin{figure}
\centering

\includegraphics[width=0.5\textwidth]{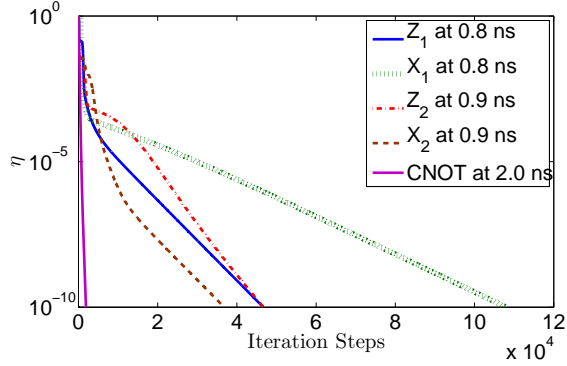}

\caption{\label{fig:error} (Color online) Gate errors $\eta$ as a function of the
  number of iterations. The gate errors calculated using Hamiltonian 
  Eq.~(\ref{eq:Hamiltonian06}) decrease as
  the number of 
  iterations increases. The stopping iteration criteria of error is
  set to be $\eta<10^{-10}$.
The single-qubit gates of the first qubit, the single-qubit gates of
the second qubit and the two-qubit CNOT gate 
 are achieved at $T=0.8$ ns, $T=0.9$ ns and $T=2.0$ ns, respectively.
}
\end{figure}

\begin{figure}
\centering

\subfigure[][]{ \label{fig:X1fc1}

\includegraphics[angle=90,width=0.45\textwidth]{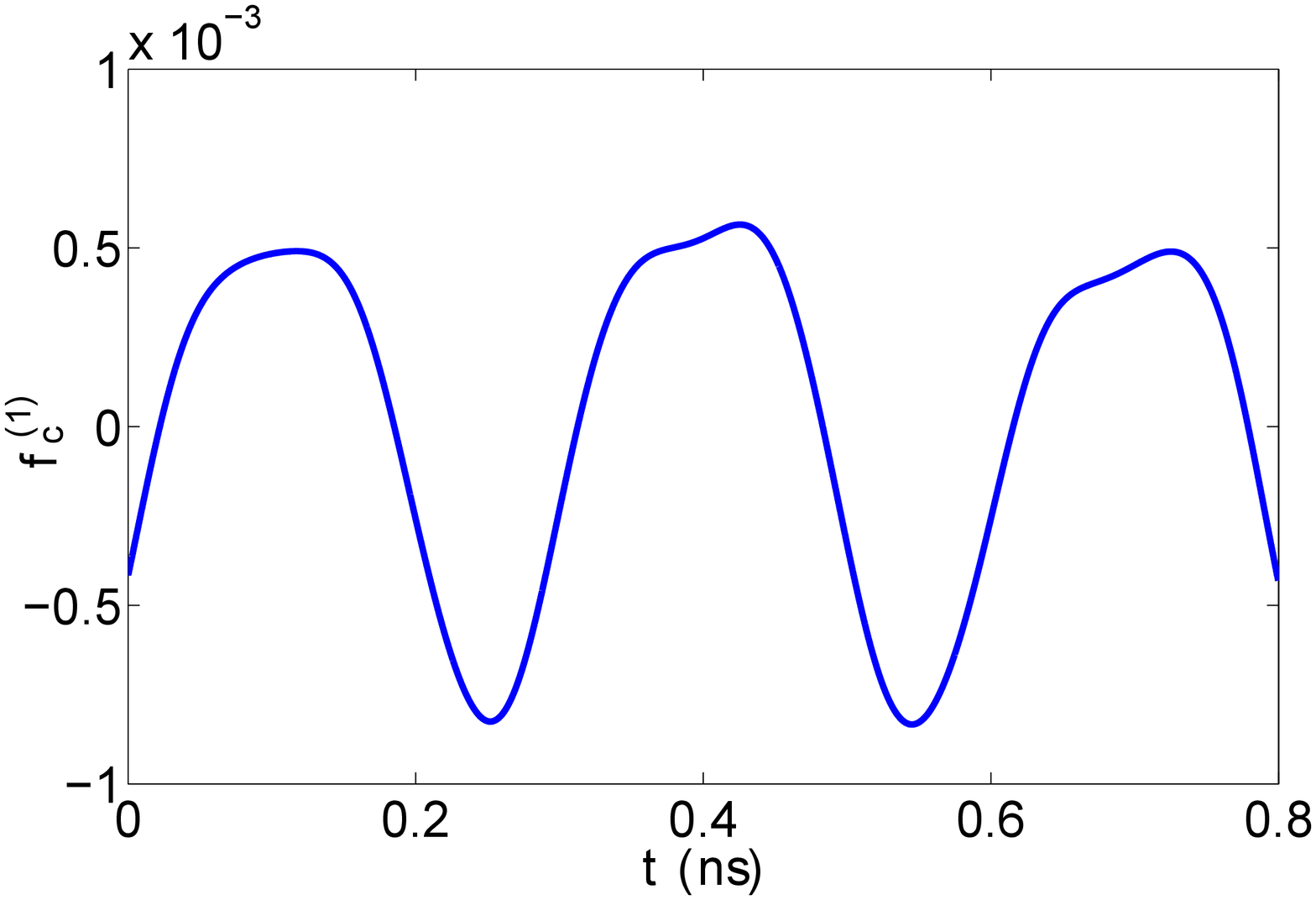}

}\subfigure[][]{ \label{fig:X1fc2}

\includegraphics[angle=90,width=0.45\textwidth]{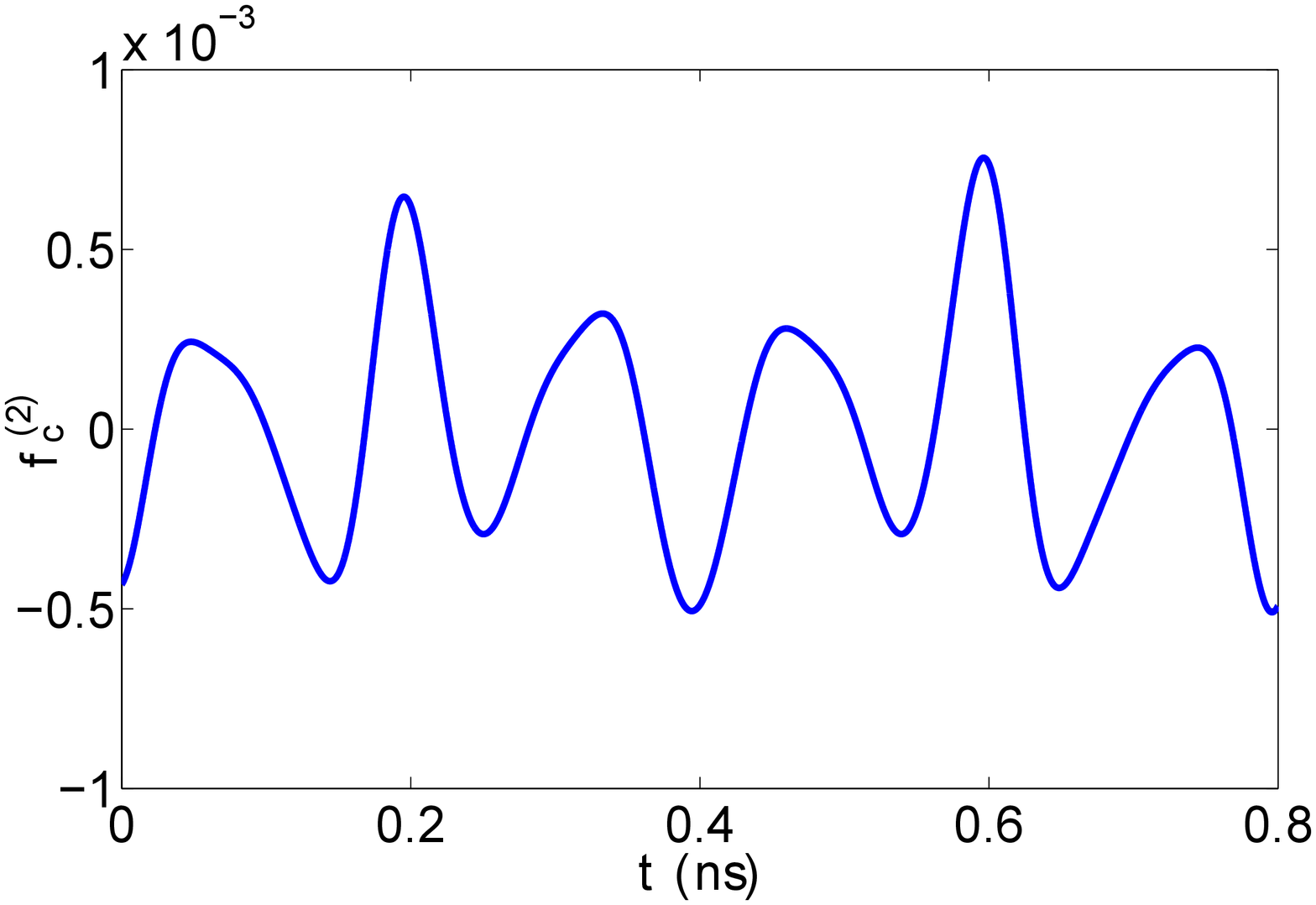}

}

\subfigure[][]{ \label{fig:X1P0}

\includegraphics[width=0.45\textwidth]{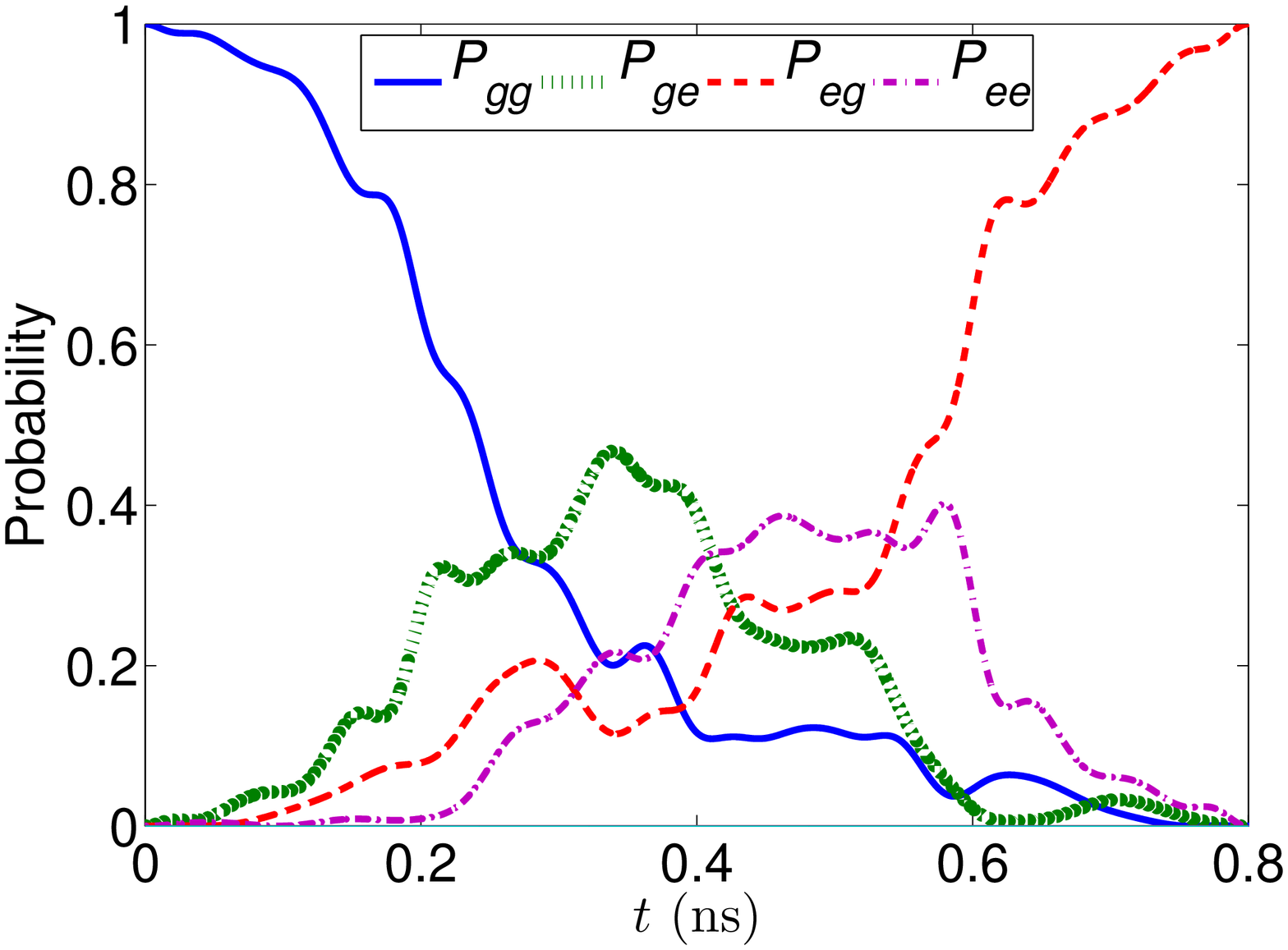}

}\subfigure[][]{ \label{fig:X1P1}

\includegraphics[width=0.45\textwidth]{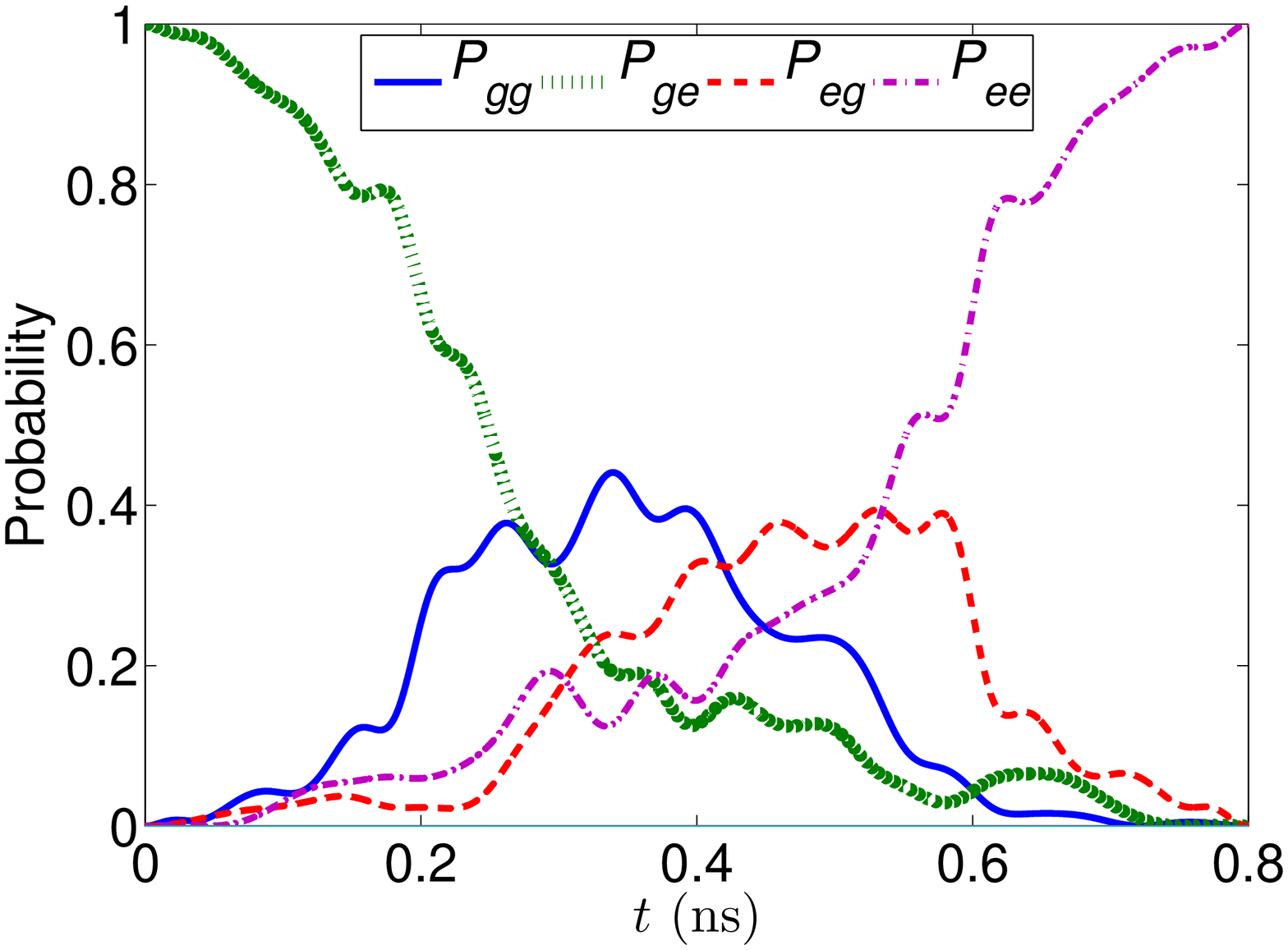}

}

\subfigure[][]{ \label{fig:X1P5}

\includegraphics[width=0.45\textwidth]{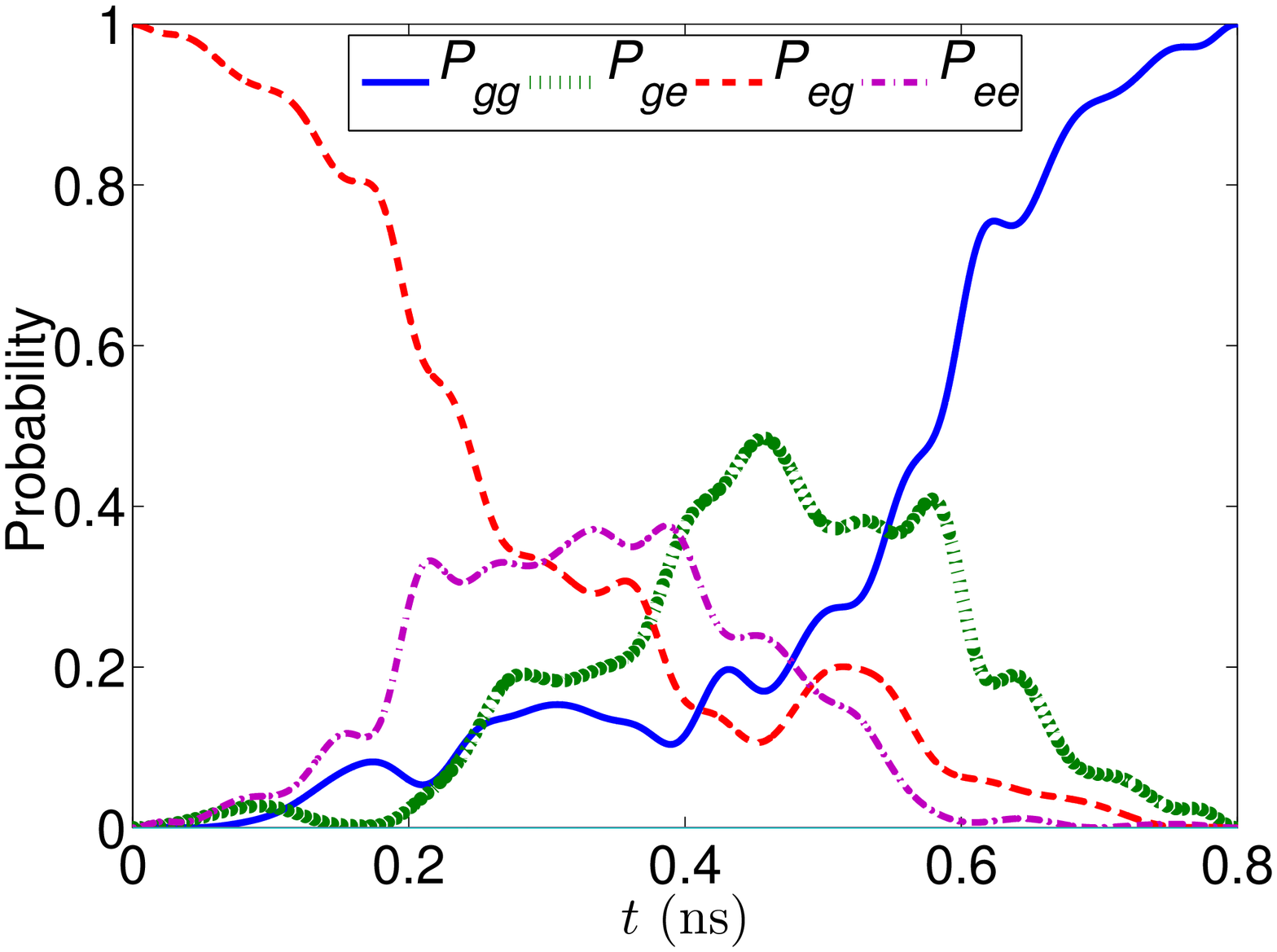}

}\subfigure[][]{ \label{fig:X1P6}

\includegraphics[width=0.45\textwidth]{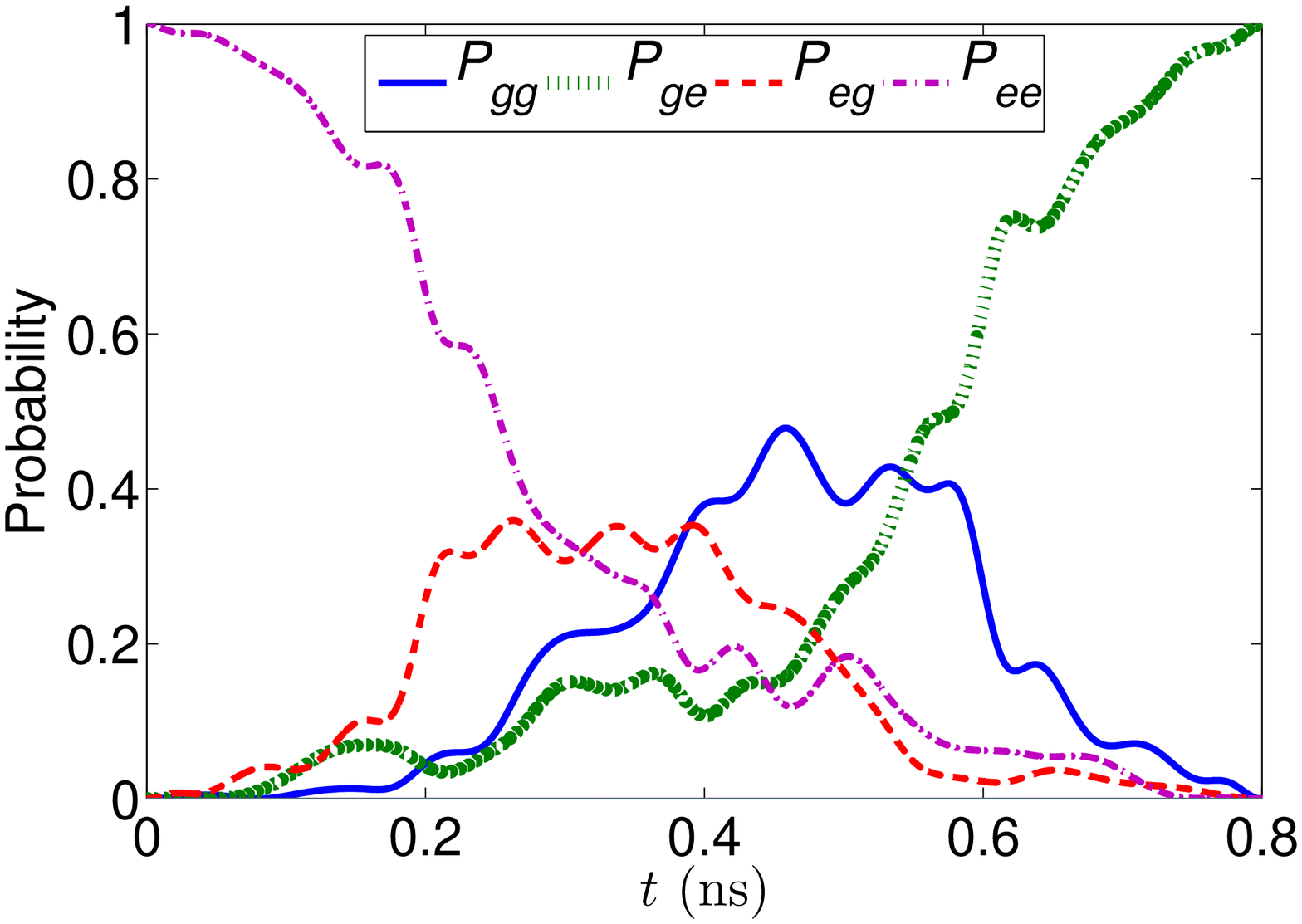}

}

\caption{\label{fig:X1} (Color online)
Optimal control sequences and state probability evolutions for a $X_1$
gate with operation time of $T=0.8$ ns.
The optimal control pulse sequences of the reduced magnetic fluxes (a)
$f_c^{(1)}(t)$ and (b) $f_c^{(2)}(t)$ are obtained via the Hamiltonian, 
Eq.~(\ref{eq:Hamiltonian05}).
The time evolutions of the two-qubit computational-state probabilities
calculated via the Hamiltonian, Eq.~(\ref{eq:Htot}), with the lowest
five energy states per qubit using the
optimal control sequences shown in (a) and (b) are plotted in (c)-(f)
with each two-qubit computational state as an initial state.
The two-qubit computational basis
states are $|gg\rangle$, $|ge\rangle$, $|eg\rangle$ and $|ee\rangle$.
}
\end{figure}

{\it Conventional approach - }
In most of the studies for the directly coupling schemes, 
the Hamiltonian of two inductively coupled flux qubits at the
optimal bias point contains only the first line of 
Eq.~(\ref{eq:Hamiltonian05}):
\begin{eqnarray}
\frac{H}{\hbar} & = &
\sum_{l=1}^{2}\left[\frac{\omega_{l}}{2}\sigma_{z}^{\left(l\right)}+\kappa_{2}^{\left(l\right)}(t)\sigma_{x}^{\left(l\right)}\right]+\Lambda_{22}\sigma_{x}^{\left(1\right)}\sigma_{x}^{\left(2\right)}.
\label{eq:Hamiltonian06}
\end{eqnarray}
This is because the coefficients of the last three time-dependent terms of
Eq.~(\ref{eq:Hamiltonian05}) are 
%very small when compared with 
 several orders of magnitude smaller than 
the qubit transition frequencies $\omega_l$, 
%qubit detuning $\left|\omega_{1}-\omega_{2}\right|$, 
the single-qubit driving amplitudes $\kappa_2^{(l)}$ and the fixed static 
qubit-qubit interaction strength $\Lambda_{22}$. Thus neglecting these coefficients does
not introduce appreciable error. Furthermore, since the strength
$\Lambda_{22}$ of 
the off-diagonal coupling $\sigma_{x}^{\left(1\right)}\sigma_{x}^{\left(2\right)}$
in the presence of large qubit detuning
$\left|\omega_{1}-\omega_{2}\right|$ affects the qubit dynamics only
in second order in the parameter 
$(\Lambda_{22}/\left|\omega_{1}-\omega_{2}\right|)$, the constant 
coupling $\Lambda_{22}$ term is often neglected at the optimal bias point 
provided that $(\Lambda_{22}/\left|\omega_{1}-\omega_{2}\right|)$ is
rather small. 
In this case, the
qubits are regarded to be effectively decoupled. Thus the single-qubit
gate, say, $X_{1}$ gate (rotation of angle $\pi$ around $x$-axis on qubit 1),  
can be achieved by simply driving the magnetic flux through the first
qubit
with microwave resonance with the transition frequency of the first qubit,
i.e., $\kappa_{2}^{(1)}=\tilde{\kappa}_{2}^{(1)}\cos(\omega_1 t)$, 
while keeping the time-dependent flux through the second qubit off, i.e.,
$\kappa_{2}^{(2)}=0$. 
When the driving field strength $\tilde{\kappa}_{2}^{(1)}$
 is much smaller 
than the transition frequencies $\omega_{l}$ and qubit detuning
$\left|\omega_{1}-\omega_{2}\right|$ in
which the rotating-wave 
approximation can be made, Eq.~(\ref{eq:Hamiltonian06}) in the frame
rotating with the driving frequency $\omega_1$ becomes
\begin{equation}
\frac{H_{{\rm rot}}}{\hbar}=\frac{\tilde{\kappa}_{2}^{\left(1\right)}}{2}\sigma_{x}^{\left(1\right)}+\frac{\omega_{2}}{2}\sigma_{z}^{\left(2\right)}.
\end{equation}
Performing single-qubit $X_1$-gate on qubit 1 will also demand qubit 2
to return
to its original state up to a global phase at the end of the $X_1$
operation, i.e., an identity gate on qubit 2. 
In this rotating frame, qubit 2 takes $T=2 n\pi/\omega_{2}$ to complete
an identity gate up to a global phase with $n$ being an integer
number. Taking this time to be the time to complete a $\pi$-pulse on qubit 1  
yields the single-qubit driving strength or the Rabi
frequency to be $\tilde{\kappa}_{2}^{\left(1\right)}=\omega_{2}/2n$.
By requiring the reduced time-dependent magnetic flux
$|f_{c}^{(1)}(t)|=|\Phi_{e}^{(1)}(t)/\Phi_{0}|\lesssim 10^{-3}$ such that
the weak-amplitude approximation to obtain
Eq.~(\ref{eq:Hamiltonian02}) is valid, one obtains the minimum integer
number $n$ to be $7$. This then leads to $f_{c}^{(1)}(t)\approx 5.77\times
10^{-4}\cos(\omega_1 t)$ and the operation time of $X_1$ gate
$T=14\pi/\omega_{2}\approx 0.85$ ns. Plugging these numbers into  
Eq.~(\ref{eq:Hamiltonian05}) and  Eq.~(\ref{eq:Htot})
to simulate an $X_1$ gate gives an error of about 
$\eta\approx\eta_P\sim3.3\times10^{-3}$.
This result indicates that the dominant source of error is mainly due
to the constant qubit-qubit 
interaction term of $\Lambda_{22}$ and does not come from the 
%effect of including higher-energy-level states 
leakage to higher-energy-level states \cite{You2005}.
Indeed, for the experimental parameters used here, the ratio of
$(\Lambda_{22}/\left|\omega_{1}-\omega_{2}\right|)=0.4/4.96\approx
0.08$, and taking the second order correction gives an estimated error 
also in the order of $10^{-3}$. This is about the best one can do in a unitary
case if neglecting the constant qubit-qubit interaction $\Lambda_{22}$ term. 
How to include the qubit-qubit interaction term and perform a much more
accurate single-qubit gate is not intuitively obvious.

{\it Optimal control approach - }
Here, we apply the Krotov
iterative method to obtain the optimal control sequences for
high-fidelity (low-error) single-qubit and two-qubit gates taking all
the static and time-dependent single-qubit and two-qubit terms or interactions
into account. 
Usually, stronger strengths of the control fields result in shorter
gate operation times. In our case,  
the gate operation times are, however, chosen such that the reduced
time-dependent magnetic fluxes satisfying the weak-amplitude
approximation of 
$|f_{c}^{(l)}(t)|=|\Phi_{e}^{(l)}(t)/\Phi_{0}|\lesssim 10^{-3}$.  
We find that the high-fidelity single-qubit gates on qubit 1
can be achieved at a gating time of $T=0.8$ ns,  and the gating time is
$T=0.9$ ns for qubit 2. The two-qubit gates will be discussed in
the next subsection.
As shown in Fig.~\ref{fig:error}, the errors $\eta$ of $Z_{1}$, $X_{1}$,
$Z_{2}$ and $X_{2}$ gates calculated using the reduced Hamiltonian, Eq.~(\ref{eq:Hamiltonian05}),
decrease with the number of iterations. 
We note here that for simplicity, we let the positive
  shape function $S(t)$ in Eq.~(\ref{eq:CostFun02}) be a
  constant. In addition, the time-dependent control field
  $\epsilon(t)$ in Eq.~(\ref{eq:CostFun02}) in our case becomes the 
reduced time-dependent magnetic fluxes
$|f_{c}^{(l)}(t)|=|\Phi_{e}^{(l)}(t)/\Phi_{0}|$ that are dimensionless
and required to be smaller than  $10^{-3}$, and thus in our
calculations we choose
  $S(t)/\lambda=10^{-10}$ (GHz)$^{-1}$ in order to make sure that the
  iterations converge to the optimal results monotonically.  
In principle, the gate error for a closed, unitary system can, via the
optimal control theory, be as
low as one wishes, limited only by the accuracy of the
two-level-approximation Hamiltonian, Eq.~(\ref{eq:Hamiltonian05}), and by 
the machine precision of the computation. Considering the accuracy of the
two-level-approximation Hamiltonian used, we set the
stopping criteria of error to be $10^{-10}$, and thus the iterations 
are terminated when error $\eta<10^{-10}$.
A typical set of the control sequences of the reduced time-dependent magnetic
fluxes $f_{c}^{(1)}(t)$ and $f_{c}^{(2)}(t)$ for the high-fidelity
$X_1$ gate is shown in Figs.~\ref{fig:X1}\subref{fig:CNOTfc1} and
\ref{fig:X1}\subref{fig:CNOTfc2}.

\subsection{Two-qubits gate: unitary case}

\begin{figure}
\centering

\subfigure[][]{ \label{fig:CNOTfc1}

\includegraphics[angle=90,width=0.45\textwidth]{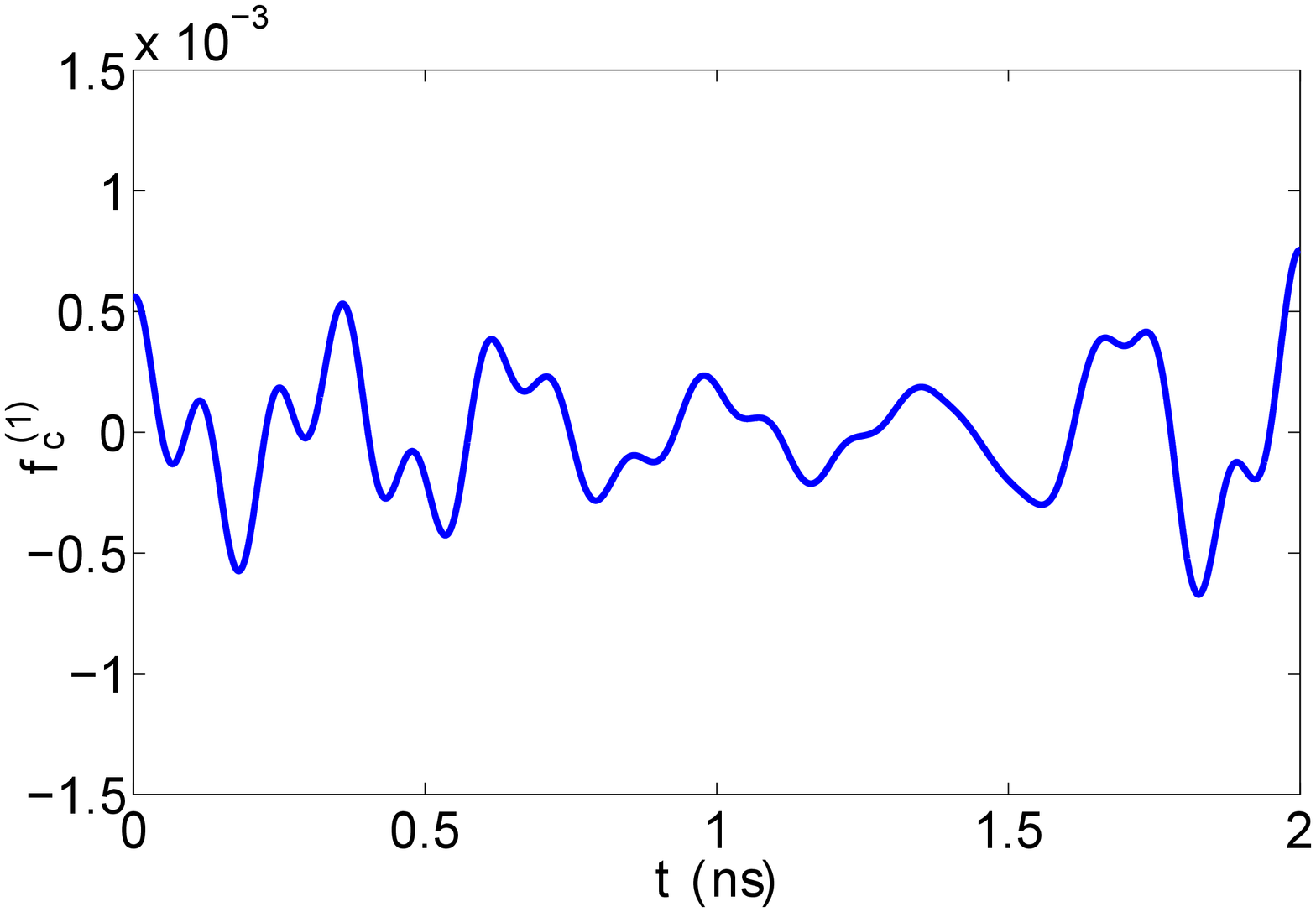}

}\subfigure[][]{ \label{fig:CNOTfc2}

\includegraphics[angle=90,width=0.45\textwidth]{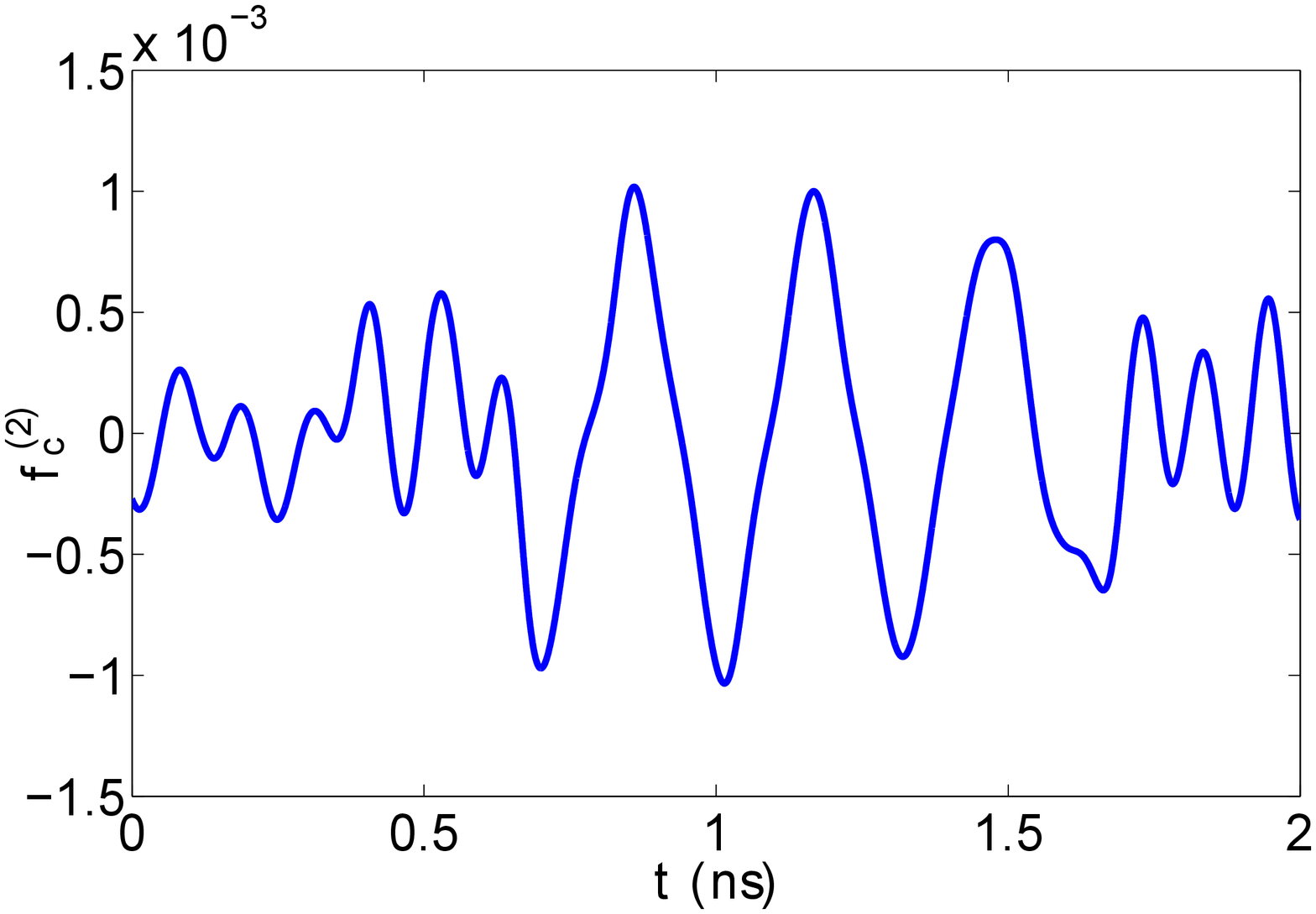}

}

\subfigure[][]{ \label{fig:CNOTP0}

\includegraphics[width=0.45\textwidth]{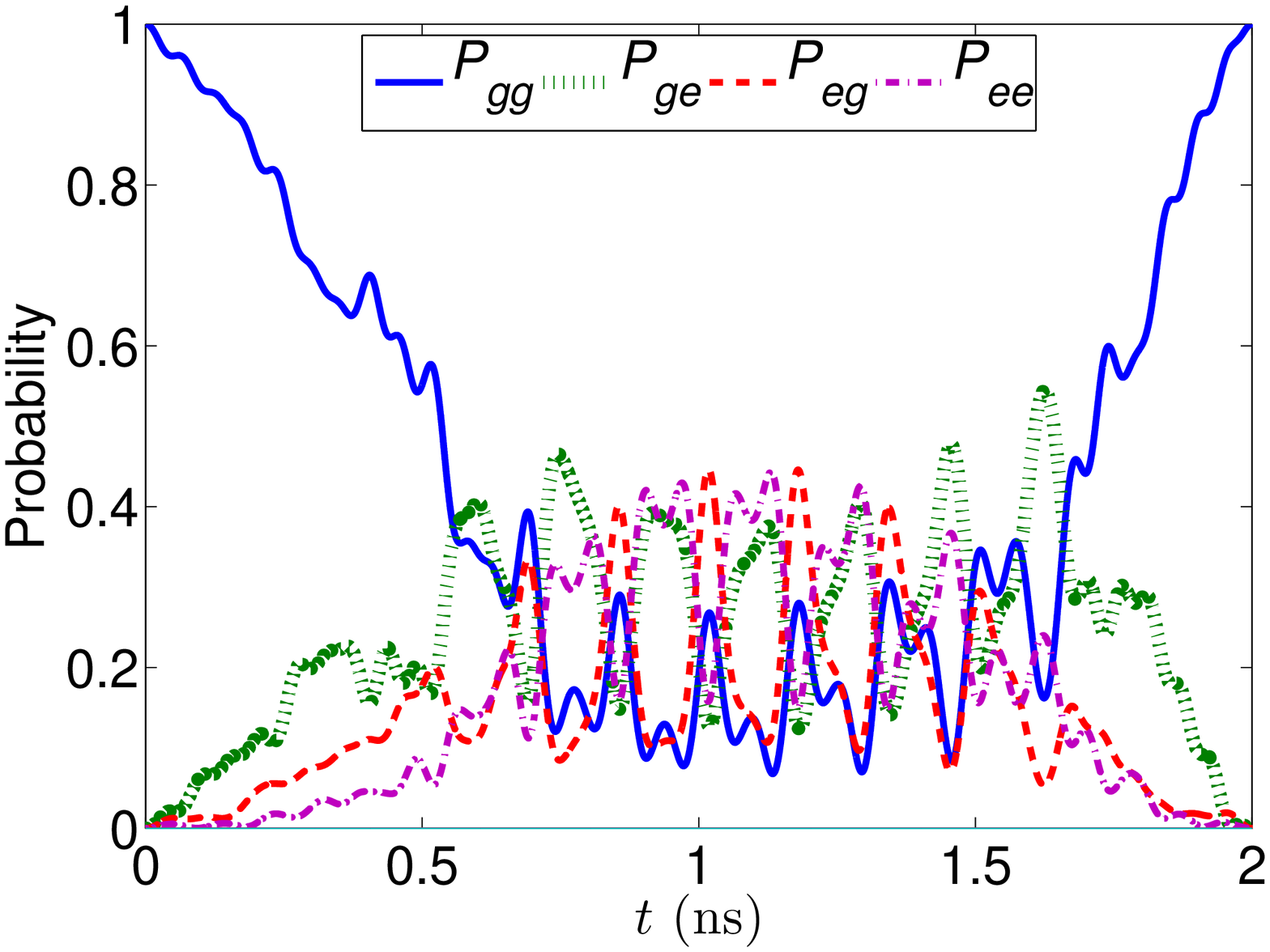}

}\subfigure[][]{ \label{fig:CNOTP1}

\includegraphics[width=0.45\textwidth]{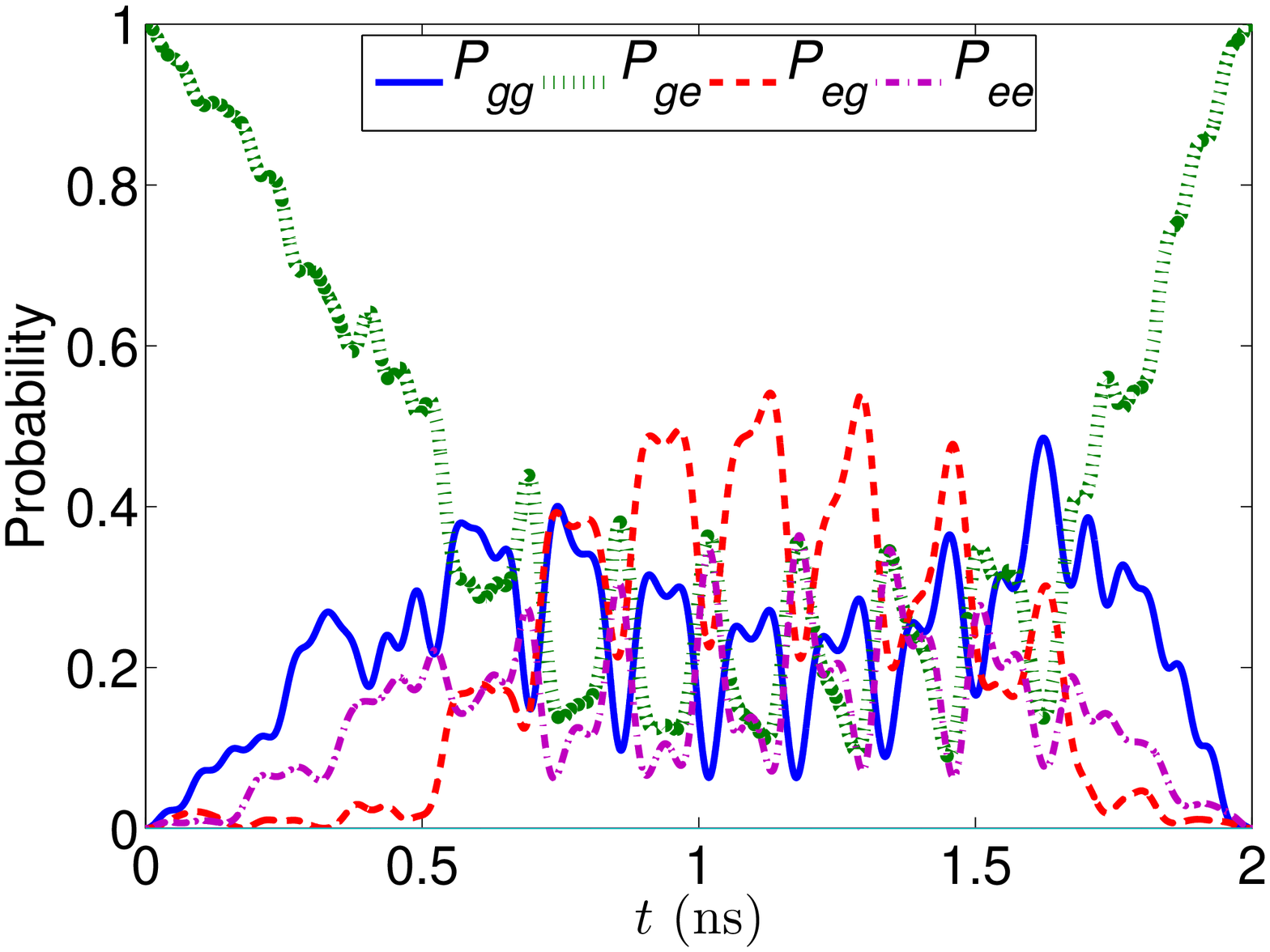}

}

\subfigure[][]{ \label{fig:CNOTP5}

\includegraphics[width=0.45\textwidth]{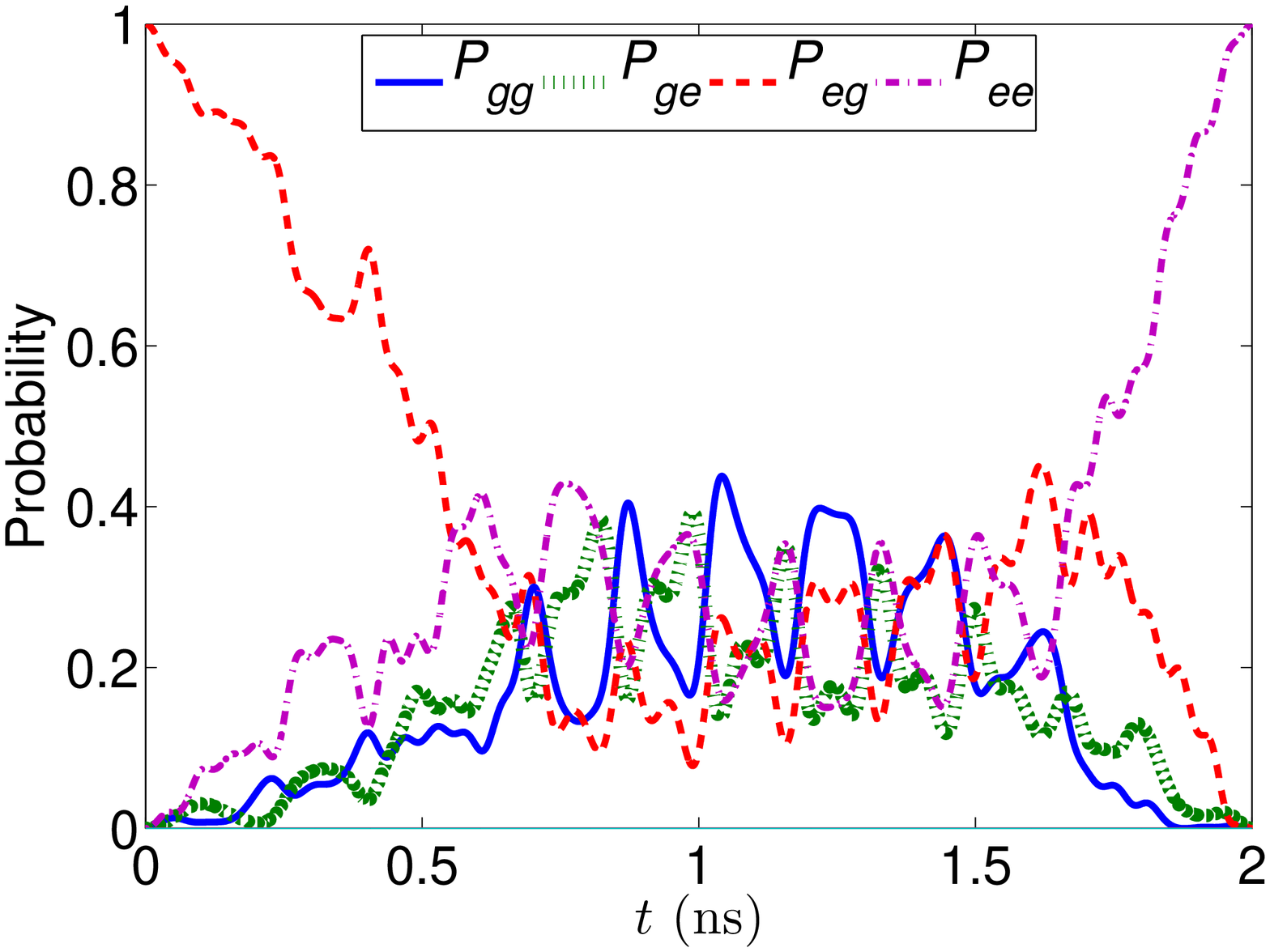}

}\subfigure[][]{ \label{fig:CNOTP6}

\includegraphics[width=0.45\textwidth]{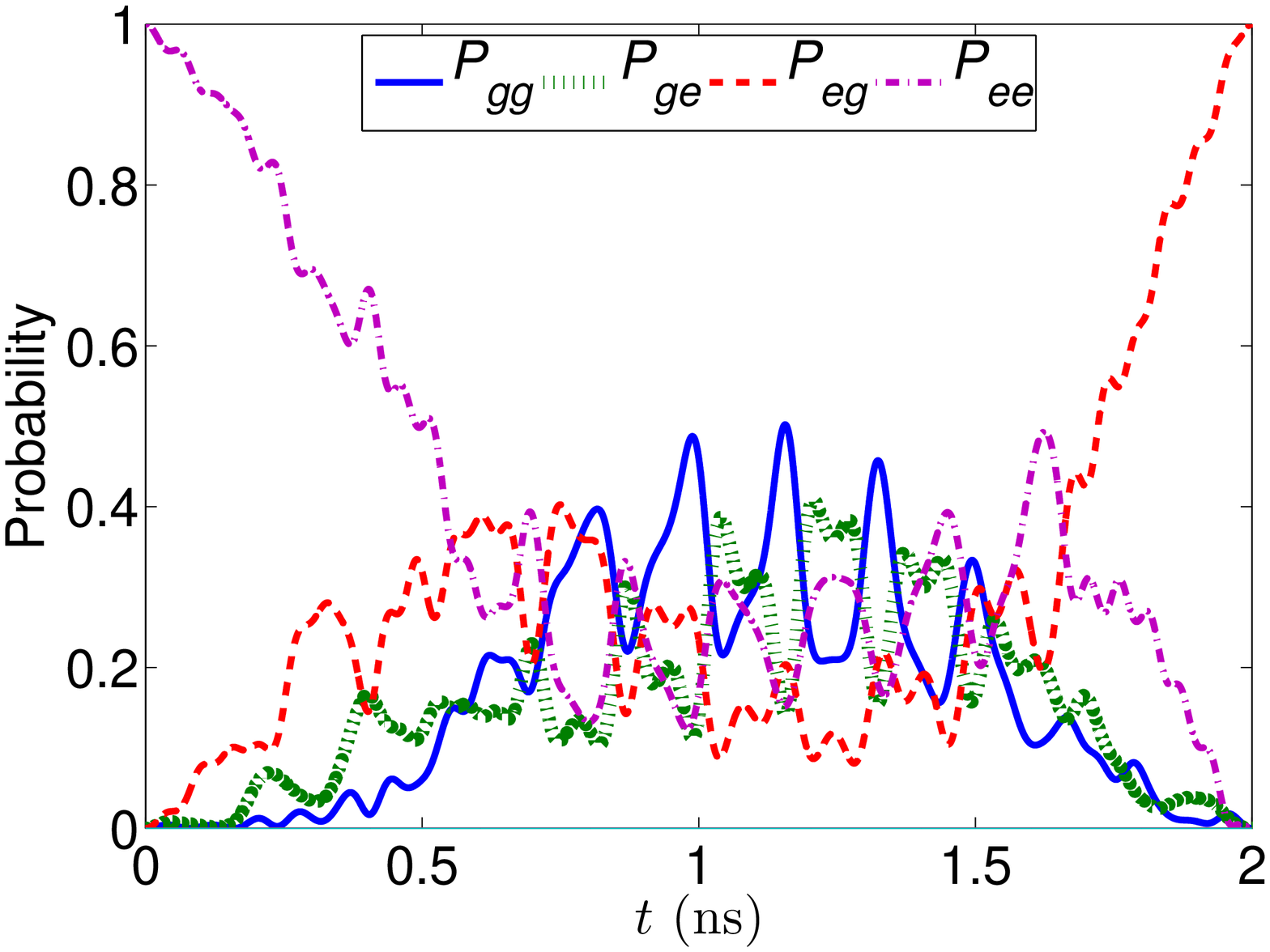}

}

\caption{\label{fig:CNOT} (Color online)
Optimal control sequences and state probability evolutions for a
two-qubit CNOT
gate with operation time of $T=2.0$ ns.
The optimal control pulse sequences of the reduced magnetic fluxes (a)
$f_c^{(1)}(t)$ and (b) $f_c^{(2)}(t)$ are obtained via the Hamiltonian, 
Eq.~(\ref{eq:Hamiltonian05}).
The time evolutions of the two-qubit computational-state probabilities
calculated via the Hamiltonian, Eq.~(\ref{eq:Htot}), with the lowest
five energy states per qubit using the
optimal control sequences shown in (a) and (b) are plotted in (c)-(f)
with each two-qubit computational state as an initial state.
The two-qubit computational basis
states are $|gg\rangle$, $|ge\rangle$, $|eg\rangle$ and $|ee\rangle$.
}
\end{figure}

{\em Conventional approach - }
Many schemes using microwaves 
have been put forward
\cite{Rigetti2005,Bertet2006,Liu06,Ashhab2006,Rigetti2010,Niskanen2006,Harrabi2009,Grajcar2006,Niskanen07}
to realize an effective controllable qubit-qubit interaction for
Hamiltonian 
(\ref{eq:Hamiltonian06}) or (\ref{eq:Hamiltonian05}).
However, in most of the schemes, two neighboring qubits are considered
effectively decoupled (as described earlier)
when no time-dependent microwave field is applied. This approximation 
limits the maximum fidelity (or minimum error) these schemes can
actually achieve. 
One of the common goals for gate control 
is to implement a two-qubit CNOT gate, an important
universal gate for quantum computation. In all of the schemes, an
entangled gate natural to their couplings is implemented, and thus
to realize a CNOT gate, an additional sequence of single-qubit rotations
is needed.  
For example, if an iSWAP gate is conveniently implemented in a
particular scheme, the CNOT gate can be realized by the following sequence
CNOT$=[I\otimes R_x(\pi/2)][R_z(-\pi/2)\otimes
R_z(\pi/2)][{\rm iSWAP}][R_x(\pi/2)\otimes I][{\rm
iSWAP}][I \otimes R_z(\pi/2)]$. 
Typical two-qubit gate operation times in these schemes are in the order of
$10$ to $200$ ns.
However, decomposing a CNOT gate or a general gate operation into
several single-qubit and some entangled two-qubit operations in series makes its operation
time normally longer than that of completing the operation in one
single run of pulses.
Moreover, the overall gate error may also become larger as the gate
errors of the 
decomposed gates will add up or accumulate.

{\em Optimal control approach - }
In contrast, the optimal control method has great advantages of
enabling the implementation of a CNOT
gate or other general quantum gates in one single run of pulse or in one single
pulse sequence by simply
setting the target 
operation to be the CNOT gate or the general quantum gate one wishes
to implement.  
Requiring the reduced time-dependent magnetic flux 
$|f_{c}^{(l)}(t)|=|\Phi_{e}^{(l)}(t)/\Phi_{0}|\lesssim 10^{-3}$, we then
set the CNOT gate operation time $T=2$ ns and use the reduced
Hamiltonian (\ref{eq:Hamiltonian05}) to find the optimal control sequence.
This CNOT gating time is about several folds to
two orders of magnitude shorter than that of the directly
or/and indirectly 
microwave controllable coupling schemes
\cite{Rigetti2005,Bertet2006,Liu06,Ashhab2006,Rigetti2010,Niskanen2006,Harrabi2009,Grajcar2006,Niskanen07}. 
This is because single-qubit and two-qubit operations can be performed
in parallel simultaneously (rather than in series) in one single run
of the optimal control pulses.
One can see from Fig.~\ref{fig:error} that the CNOT gate error drops
quickly with the number of iterations. 
Here, qubit 1 is the control qubit and qubit 2 is
the target qubit. 
The optimal control sequences obtained when
the iteration terminates at the point when error $\eta<10^{-10}$ are shown 
in Figs.~\ref{fig:CNOT}\subref{fig:CNOTfc1} and
~\ref{fig:CNOT}\subref{fig:CNOTfc2}.

\subsection{Effect of leakage states}

To investigate the gate error due to the leakage to higher
energy-level states outside the computational basis-state space, 
we apply the control sequences of the
single-qubit and two-qubit gates obtained
from the optimization of the reduced two-level-approximation
Hamiltonian (\ref{eq:Hamiltonian05}) 
to the multi-level Hamiltonian (\ref{eq:Htot}) and calculate the
errors $\eta_P$ of Eq.~(\ref{errorP})
in the projected two-qubit computational state subspace. 
The single-qubit gate errors are found to be $\eta_P=6.56\times10^{-8}$, $6.21\times10^{-8}$,
$3.25\times10^{-8}$, and $5.79\times10^{-8}$
for the $Z_{1}$, $X_{1}$, $Z_{2}$ and $X_{2}$ gates, respectively.
These results indicate that the excitations to higher energy-level
states increase slightly the gate errors; however, the gate errors in the order
of $10^{-8}$ are still much smaller than $10^{-3}$ of those  
conventional approaches that apply directly a microwave $\pi$ pulse by
neglecting the qubit-qubit interaction.  
We also simulate the $X_1$ gate numerically with the multi-level
Hamiltonian (\ref{eq:Htot}) using the optimal control sequences obtained  
from the reduced Hamiltonian (\ref{eq:Hamiltonian05})
for initial four different qubit computational basis
states $|gg\rangle$, $|ge\rangle$, $|eg\rangle$ and $|ee\rangle$.
The time evolutions of the
state probabilities of the $X_1$ gate are shown 
in Figs.~\ref{fig:X1}\subref{fig:X1P0}-\ref{fig:X1}\subref{fig:X1P6}.
As expected, after the $X_1$ operation, $|gg\rangle \to |eg\rangle$,
$|ge\rangle \to |ee\rangle$, $|eg\rangle \to |gg\rangle$, and 
$|ee\rangle \to |ge\rangle$. 
It seems that there is an approximate symmetry under time-reversal
between Figs.~\ref{fig:X1}\subref{fig:X1P0} and
\ref{fig:X1}\subref{fig:X1P5}, and between \ref{fig:X1}\subref{fig:X1P1}
and  \ref{fig:X1}\subref{fig:X1P6}. This is mainly due
to the fact that the control pulse sequences for the $X_1$ gate 
shown in Figs.~\ref{fig:X1}\subref{fig:CNOTfc1} 
and \ref{fig:X1}\subref{fig:CNOTfc2} are 
seemingly symmetric with respect to the midpoint of the operation time.
Thus the time evolutions of the state probabilities, when the
first qubit making the transition between $|g\rangle \rightarrow |e\rangle$
and $|e\rangle \rightarrow |g\rangle$ with the second qubit returning
back to the same state, appear to have such a time-reversal symmetry.
In our case, the pulse sequences are
obtained when the gate error is smaller than  
the error threshold of $\eta<10^{-10}$ for a certain initial guess of the
pulse sequence. 
The truly global optimal pulses should, however, correspond to the
minimum value of the cost function obtained by trying all possible
initial guesses of the pulse shapes. 
Thus other initial guesses can generate slightly
different $X_1$-gate pulse sequences 
satisfying still the same error threshold requirement but with less symmetry. 
In fact, the pulse sequence of $f_{c}^{(2)}(t)$ for the $X_2$ gate we
obtain (not shown here) does not have an apparently symmetric shape
with respect to the 
midpoint of the operation time as compared to  $f_{c}^{(1)}(t)$ of the $X_1$
gate in Figs.~\ref{fig:X1}\subref{fig:CNOTfc1}. 
As a result,  the
time evolutions of the state
probability between $|gg\rangle \rightarrow 
|ge\rangle$  ($|eg\rangle \rightarrow |ee\rangle$) and 
$|ge\rangle \rightarrow |gg\rangle$  ($|ee\rangle \rightarrow
|eg\rangle$) for the $X_2$ gate (not shown here) do not show prominent 
time reversal symmetry as their corresponding counterparts 
for the $X_1$ gate. 
There is, however, indeed an approximate symmetry in time evolutions
of the state probabilities in
Figs.~\ref{fig:X1}\subref{fig:X1P0}-\ref{fig:X1}\subref{fig:X1P6}. 
This approximate symmetry could be understood in terms of Hamiltonian
Eq.~(\ref{eq:Hamiltonian05}), or more easily in terms of Hamiltonian
Eq.~(\ref{eq:Hamiltonian06}) by neglecting the tiny contributions of
the last three time-dependent terms of
Eq.~(\ref{eq:Hamiltonian05}). If one puts aside the fixed static
qubit-qubit interaction $\Lambda_{22}$ term, then the two qubits in
Hamiltonian Eq.~(\ref{eq:Hamiltonian06}) are decoupled and have the
symmetry  of exchanging $|e_l\rangle$ and $|g_l\rangle$ in the presence
of $\sigma_{x}^{(l)}$ term. As a result, the time evolutions of the
state probabilities $P_{eg}$, $P_{ee}$, $P_{gg}$ and $P_{ge}$ in
Figs.~\ref{fig:X1}\subref{fig:X1P0}-\ref{fig:X1}\subref{fig:X1P6},
%Figs.~ \ref{fig:X1P0}, \ref{fig:X1P1}, \ref{fig:X1P5} and
%\ref{fig:X1P6}, 
respectively, behave approximately the same.
Similarly, the approximately same time evolutions for the state
probabilities $P_{gg}$, $P_{ge}$, $P_{eg}$ and $P_{ee}$ in 
Figs.~\ref{fig:X1}\subref{fig:X1P0}-\ref{fig:X1}\subref{fig:X1P6},
%Figs.~\ref{fig:X1P0}, \ref{fig:X1P1}, \ref{fig:X1P5} and \ref{fig:X1P6},
respectively, are also observed.

Applying the optimal control sequences of two-qubit CNOT gate shown 
in Figs.~\ref{fig:CNOT}\subref{fig:CNOTfc1} and
\subref{fig:CNOTfc2} to  multi-level Hamiltonian Eq.~(\ref{eq:Htot}) with
initial four different qubit computational basis
states $|gg\rangle$, $|ge\rangle$, $|eg\rangle$ and $|ee\rangle$, we
plot the time evolutions of the
state probabilities of the CNOT gate in 
Figs.~\ref{fig:CNOT}\subref{fig:CNOTP0}-\ref{fig:CNOT}\subref{fig:CNOTP6}.  
When qubit 1 is initially in the state $\left|g\right\rangle $,
the two qubits return to their original states; when qubit 1 is
initially in the state $\left|e\right\rangle $, an effective NOT operation is
performed on qubit 2 (the target qubit) at the end of the CNOT gate
operation. This is exactly what we see in 
Figs.~\ref{fig:CNOT}\subref{fig:CNOTP0}-\ref{fig:CNOT}\subref{fig:CNOTP6}. 
The error of the CNOT gate when including the higher energy-level states
is $\eta_P\approx 1.57\times10^{-6}$. By Comparing this error to the error of the
single-qubit gates of $\eta_P<7 \times10^{-8}$, leakage to the
higher-energy level states outside the computational basis state space is
more appreciable for 
the CNOT gate control pulses. This may be due to the fact that the
magnitude of the reduced time-dependent magnetic
flux $f_{c}^{(2)}(t)$ of 
Fig.~\ref{fig:CNOT}\subref{fig:CNOTfc2} for the CNOT gate is larger
than  $f_{c}^{(1)}(t)$ and $f_{c}^{(2)}(t)$ for the 
single-qubit gates [see, e.g., $f_{c}^{(1)}(t)$ and $f_{c}^{(2)}(t)$
of Figs.~\ref{fig:X1}\subref{fig:CNOTfc1} and 
\ref{fig:X1}\subref{fig:CNOTfc2} for the $X_1$ gate].
Nevertheless, the CNOT gate error of the two coupled flux qubits is, to our
knowledge, still much 
lower than the currently available schemes when the leakage to
higher energy-level states is considered.  
We have also performed an optimal CNOT gate operation with qubit 1 being the
target qubit and qubit 2 being the control qubit. This CNOT gate can
also be accomplished at $T=2.0$ ns with error $\eta_P\approx 6.60\times10^{-7}$.

\subsection{Effect of decoherence}
We show below that our optimal control scheme is robust against
qubit decoherence. We use the Born-Markov master equation for the reduced
density matrix of the flux qubit system in the Lindblad form to model
the effect of decoherence on the qubit dynamics \cite{Carmichael99,Gardiner00,Breuer02}:
\begin{eqnarray}
\dot{\rho}(t) 
 & = &
 -\imath\left[H,\rho(t)\right]+\sum_{i=1}^{2}\left(\Gamma_{1,i}\mathcal{D}\left[\sigma_{i}^{-}\right]\rho(t)+\Gamma_{\varphi,i}\mathcal{D}\left[\sigma_{i}^{z}\right]\rho(t)\right),
\label{eq:masterEq1}\\
&\equiv& \mathcal{L}\rho(t),
\label{eq:masterEq}
\end{eqnarray}
where the Hamiltonian $H$ is defined in Eq.~(\ref{eq:Hamiltonian05}),
the superoperator
\begin{equation}
\mathcal{D}\left[c\right]\rho=c\rho
c^{\dagger}-\frac{1}{2}c^{\dagger}c\rho-\frac{1}{2}\rho c^{\dagger}c, 
\end{equation}
and $\Gamma_{1,i}$ and $\Gamma_{\varphi,i}$ are the relaxation rate and
dephasing rate of qubit $i$, respectively. 
Generally speaking, if one starts
 from a microscopic model in the weak system-environment coupling
 limit, the argument in
 the superoperator $\mathcal{D}$ in the Born-Markov approximation
should rather be operators generating transitions
 between the energy eigenstates of the coupled-qubit Hamiltonian
 (\ref{eq:Hamiltonian05}) \cite{Breuer02}. 
However, for small inter-qubit coupling considered here, these
operators in the rotating-wave approximation can be
 approximately represented by individual qubit transition operators,
 $\sigma_i^-$ and $\sigma_i^+$. In addition,  at
 finite temperatures one should have a dissipative term describing the influx of energy
 from the thermal environment 
into the system. But since typical flux-qubit experiments are
 conducted at a low temperature around 50 mK \cite{Bylander11} which
 is much smaller 
 than the transition frequencies of the qubits considered here (i.e.,
 $\hbar\omega_{l}/k_{B}T\gg 1$). As a result, the thermal mean
 occupation number of the environment modes at the energy of about
 $\hbar\omega_{l}$ approaches zero, and the environment
 may be regarded as an effective zero-temperature bath. These justify
 the use of the master equation (\ref{eq:masterEq1}) to
 phenomenologically model the
 decoherence dynamics of the coupled flux-qubit system.    
The master equation is symbolically expressed in a concise form with
a Liouville's superoperator $\mathcal{L}$ in 
%the second line of 
Eq.~(\ref{eq:masterEq}). The equation of motion of the propagator can
be found, by substituting  $\rho(t)=\mathcal{G}\left(t\right)\rho(0)$ into 
Eq.~(\ref{eq:masterEq}), to be 
\begin{equation}
\dot{\mathcal{G}}\left(t\right)=\mathcal{L}\mathcal{G}\left(t\right).
\label{eq:G}
\end{equation}
With the relaxation time $T_{1}=13\,\mu s$
obtained from the Rabi oscillation and the decoherence time $T_{2}=2.5\,\mu s$
from Ramsey interference measurements in an recent experiment
\cite{Bylander11} that
probed the noise spectrum of a superconducting flux qubit, one is able to
deduce the realistic values of $\Gamma_{1,i}$ and $\Gamma_{\varphi,i}$ using
the relations of $\Gamma_{2,i}=\Gamma_{1,i}/2+\Gamma_{\varphi,i}$,
$\Gamma_{1,i}=1/T_{1,i}$ and $\Gamma_{2,i}=1/T_{2,i}$.  
Define the gate error $\eta_{D}$  by replacing   
 $U\left(T\right)$ with $\mathcal{G}\left(T\right)$
in Eq.~(\ref{error1}) for the case considering qubit
decoherence. One can perform the optimal control
calculation by minimizing the cost function Eq.~(\ref{eq:CostFun02})
with $\eta \to \eta_D$ using the equation of motion (\ref{eq:G}). 
The single-qubit gate errors considering the effect of decoherence 
are found to be in the order of $10^{-6}$, and the errors are in the order
of $10^{-5}$ for two-qubit CNOT gates. These gate errors are still
below the error threshold $10^{-4}$ 
($10^{-3}$ in \cite{Aliferis2009}; $10^{-2}$ if surface code error
correction is used \cite{Wang11,Fowler12L,Fowler12}) 
required for fault-tolerant quantum computation. 
We have also tested numerically that using a finite-temperature
master equation with arguments in
 the superoperator $\mathcal{D}$ 
being the operators generating transitions
 between the energy eigenstates but with the decay rates and the
 dephasing rates assuming to be the same as individual single qubits
leads to the similar error values with correction only to the digits in
$10^{-8}$ or  $10^{-7}$.   
%We calculate also the gate errors by applying the optimal control pulses
%obtained from the dissipative Liouville's equation (\ref{eq:G}) to the multi-le%vel Hamiltonian, Eq.~(\ref{eq:Htot}), the
%results of the gate errors are similar to those of using ideal unitary
%pulses obtained from the reduced Hamiltonian (\ref{eq:Hamiltonian05}),
%i.e., leakage to higher-energy-level states is not appreciable. This
%may due to the fact that our gate operation times are all much smaller than
%the qubit decoherence time 
%so that at the end of the gate operations the optimal control pulse sequences
%obtained for the dissipative case do not deviate much from those for
%the unitary case. 

\section{Discussion and Conclusion}
\label{sec:Conclusion}

\begin{table}
\centering
\caption{\label{tab:gate_error}Summary of the quantum gate errors}
\begin{tabular}{|c|c|c|c|c|}
\hline 
gate & gate time & $\eta$ (unitary)& $\eta_{P}$ (leakage)& $\eta_{D}$
(decoherence) \tabularnewline
\hline
\hline 
$Z_{1}$ & $0.8$ ns & $<10^{-10}$ & $6.56\times10^{-8}$ & $2.46\times10^{-6}$ \tabularnewline
\hline 
$X_{1}$ & $0.8$ ns & $<10^{-10}$ & $6.21\times10^{-8}$ &
$2.34\times10^{-6}$ \tabularnewline
\hline 
$Z_{2}$ & $0.9$ ns & $<10^{-10}$ & $3.25\times10^{-8}$ & $3.84\times10^{-6}$ \tabularnewline
\hline 
$X_{2}$ & $0.9$ ns & $<10^{-10}$ & $5.79\times10^{-8}$ & $3.76\times10^{-6}$ \tabularnewline
\hline 
CNOT$_{1}$ & $2.0$ ns & $<10^{-10}$ & $6.6\times10^{-7}$ &
$1.42\times10^{-5}$ \tabularnewline
\hline 
CNOT$_{2}$ & $2.0$ ns & $<10^{-10}$ & $1.57\times10^{-6}$ & $1.42\times10^{-5}$ \tabularnewline
\hline 
\end{tabular}
\end{table}

We have applied the quantum optimal control method to  
implement fast and high-fidelity single-qubit and two-qubit gates for two
inductively coupled superconducting flux qubits with fixed static off-diagonal
qubit-qubit coupling and fixed qubit transition frequencies. 
Table \ref{tab:gate_error} summarizes the gate errors calculated with
realistic experimental parameters for
ideal unitary case, and for the cases considering the effect of leakage
state and the effect of qubit decoherence. 
Our optimal control scheme has the following great advantages.
(1) Our scheme that shares the same advantage of other directly
coupling scheme requires no additional coupler subcircuit and control
lines and thus is simple in experimental design.  
The control lines needed are only for the manipulation of individual qubits 
(e.g., the time-dependent magnetic flux or field through each qubit). 
(2) Quantum gates constructed via our scheme are all
with very high fidelity 
(very low error) as our optimal control scheme takes into account the
fixed qubit detuning and fixed two-qubit
interaction as well as all other time-dependent magnetic-field-induced
single-qubit interactions and two-qubit couplings  when 
performing single-qubit and two-qubit gates.
(3) Our scheme can cope with noise and decoherence very well as the
qubits are biased at the optimal coherence point to reduce the influence
of low-frequency flux noise, and the gate operation time ($\sim 2$
ns) is
much shorter than the corresponding qubit decoherence time
(a few $\mu$s or larger) \cite{Bylander11}
%(hundreds of ns to a few $\mu$s or larger)
%\cite{Clarke2008,Yoshihara2006,Kakuyanagi2007,Chow2010,Paik2011,Corcoles2011,Rigetti2012}
so that the decoherence effect on such a fast gate is significantly diminished. 
(4) A CNOT gate or other general quantum gates can be
implemented in a single run of pulse sequence rather than being
decomposed into several single-qubit and some entangled two-qubit
operations in series by composite pulse sequences.

A natural question then arises: how to generate these optimal control
pulses experimentally? The pulse sequences shown 
in Fig.~\ref{fig:X1}\subref{fig:CNOTfc1} and
\subref{fig:CNOTfc2}, and in
 Fig.~\ref{fig:CNOT}\subref{fig:CNOTfc1} and
\subref{fig:CNOTfc2}
look experimentally challenging, but not impossible. 
Commercial devices (e.g., Tektronix AWG70001A) for generating
arbitrary wave forms with 10 bits of vertical resolution at a sample
rate of 50 GSa/s, a bit rate of 12.5 Gb/s and a rise/fall time smaller
than 27 ps are now available. Such a device should enable generation of
complex signals in a time scale of sub-nanoseconds to nanoseconds.
This high-end arbitrary wave form generator 
combining with ultrafast Josephson electronics
should be or about to be able to fulfill the necessary specifications
for implementing 
the optimal control pulse sequences we obtain for the quantum gate
operations presented here.

Another challenge for the experimental implementations and applications
of the quantum optimal control theory is one's imprecise knowledge of the 
quantum system's parameters. 
%Hamiltonian. 
Typically, the quantum gates constructed by the quantum optimal
control theory are computed to very 
high precision assuming the parameters in the model Hamiltonian are
exactly known. However, for real systems controlled in the
experiments, the parameters and also the Hamiltonians usually are not
known exactly. 
%, and the Hamiltonians for the system environment coupling are known
%to an even lesser degree. 
This poses a challenge to implement such high-fidelity gates
successfully in the
laboratory.  Recently, a hybrid open/closed-loop optimal control
method called adaptation by hybrid optimal control (Ad-HOC) has been
proposed \cite{Egger14} to overcome not only the problem of inaccurate
knowledge of the 
system parameters but also
shortcomings of 
the assumed physical model and
errors on the controls themselves.
The basic idea is to use the open-loop  quantum optimal control theory
to find the optimal control pulses with the
best available model and parameters of the system.  
Then the pulses are 
sent to the experiment and their performance are efficiently measured
by using, for example, the method of randomized benchmarking
\cite{Magesan2012,Chow2009,Kelly14}.  
The closed-loop pulse calibration of Ad-HOC, similar to adaptive model-free feedback control (also
referred to as closed-loop laboratory control or learning control) \cite{Rabitz1992,Rabitz2010},
uses the physical system itself as a feedback to calibrate 
control pulses and optimize their performance.
By using, e.g., the robust and efficient
Nelder-Mead algorithm \cite{Nelder1967,Egger14,Kelly14}, the control
pulses are updated experimentally and the
procedure is iterated until a target performance is reached or
convergence stops. 
%This forms the Ad-HOC closed-loop optimization of the experimental system. 
After the calibration is finished, the optimized pulses can then be used. 
Even though the precise experimental parameters are never
identified, Ad-HOC can reduce the initial gate errors of the numerically
obtained optimal control pulses implemented in an
experiment by at least an order of magnitude \cite{Egger14}. 
Thus the quantum optimal control theory is practical and applicable experimentally through, e.g., 
the Ad-HOC protocol \cite{Egger14}. 
%use the model parameters drawn from the error bars of the
%initial measurement or characterization of the system 
% and imprecise knowledge of the
%system parameters. 
A recent experiment demonstrating single-qubit
operations (with fidelity of 0.99) 
and two-qubit entanglement for the electron spins of two proximal
nitrogen-vacancy centers in diamond using
optimal control has been reported \cite{Dolde13}.
%{\color{red} However, the systematic error, which is around $10$ MHz in the typical system, like imprecise characterization of parameters entering the Hamiltonian, is not counted. In order to overcome the errors on the controls themselves and inaccurate knowledge of the parameters, the technique called Adaptation by hybrid optimal control (Ad-HOC) can be applied \cite{Egger14}.}
It is thus believed that high-fidelity quantum gates with optimal control pulses
obtained via our scheme will be realized experimentally in the
near future.

\begin{acknowledgments}
We are very grateful to Prof. Yu-xi Liu for sending us his note for
the derivation of the coupled flux-qubit Hamiltonian of Ref.~\cite{Liu06}.  
We acknowledge support from the
National Science Council in Taiwan under Grant
No.~100-2112-M-002-003-MY3, 
from the National Taiwan University under Grants
No.~103R891400, No.~103R891402 and 102R3253, and
from the
focus group program of the National Center for Theoretical
Sciences, Taiwan.
\end{acknowledgments}

%\section*{References}


\begin{thebibliography}{99}
\bibitem{You2005P} J. Q. You and F. Nori, Phys. Today {\bf 58}, 11, 42 (2005); J. Q. You and F. Nori, Nature {\bf 474}, 589
  (2011), and references therein.
\bibitem{Clarke2008} J. Clarke and F. K. Wilhelm, Nature {\bf 453},
  1031 (2008), and references therein.
%\bibitem{You2011} J. Q. You and F. Nori, Nature {\bf 474}, 589 (2011), and references therein. 
\bibitem{Nakamura1999} Y. Nakamura, Yu. A. Pashkin, and J. S. Tsai, Nature {\bf 398}, 786 (1999).
\bibitem{Vion2002} D. Vion, A. Aassime, A. Cottet, P. Joyez, H. Pothier, C. Urbina, D. Esteve, and M. H. Devoret, Science {\bf 296}, 886 (2002).
\bibitem{Chiorescu2003} I. Chiorescu, Y. Nakamura, C. J. P. M. Harmans, and J. E. Mooij, Science {\bf 299}, 1869 (2003).

\bibitem{Bertet2005}P. Bertet, I. Chiorescu, G. Burkard, K. Semba, C. J. P. M. Harmans, D. P. DiVincenzo, and J. E. Mooij, Phys. Rev. Lett. {\bf 95}, 257002 (2005).

\bibitem{Wallraff2005} A. Wallraff, D. I. Schuster, A. Blais, L. Frunzio, J. Majer, M. H. Devoret, S. M. Girvin, and R. J. Schoelkopf, Phys. Rev. Lett. {\bf 95}, 060501 (2005).
\bibitem{Martinis2005} J. M. Martinis, K. B. Cooper, R. McDermott, M. Steffen, M. Ansmann, K. D. Osborn, K. Cicak, S. Oh, D. P. Pappas, R. W. Simmonds, and C. C. Yu, Phys. Rev. Lett. {\bf 95}, 210503 (2005).


\bibitem{Yoshihara2006}F. Yoshihara, K. Harrabi, A. O. Niskanen, Y. Nakamura, and J. S. Tsai, Phys. Rev. Lett. {\bf 97}, 167001 (2006).

\bibitem{Kakuyanagi2007}K. Kakuyanagi, T. Meno, S. Saito, H. Nakano,
  K. Semba, H. Takayanagi, F. Deppe, and A. Shnirman, Phys. Rev. Lett. {\bf 98}, 047004 (2007).

\bibitem{Corcoles2011} A. D. C\'orcoles, J. M. Chow, J. M. Gambetta,
  C. Rigetti, J. R. Rozen, G. A. Keefe, M. B. Rothwell, M. B. Ketchen, and M. Steffen, Appl. Phys. Lett. {\bf 99}, 181906 (2011).

\bibitem{Rigetti2012}C. Rigetti, J. M. Gambetta, S. Poletto, B. L. T. Plourde, J. M. Chow, A. D. C\'orcoles, J. A. Smolin, S. T. Merkel, J. R. Rozen, G. A. Keefe, M. B. Rothwell, M. B. Ketchen, and M. Steffen, Phys. Rev. B {\bf 86}, 100506 (2012).



\bibitem{Chow2010} J. M. Chow, L. DiCarlo, J. M. Gambetta, F. Motzoi, L. Frunzio, S. M. Girvin, and R. J. Schoelkopf, Phys. Rev. A {\bf 82}, 040305(R) (2010).
\bibitem{Paik2011} H. Paik, D. I. Schuster, L. S. Bishop, G. Kirchmair, G. Catelani, A. P. Sears, B. R. Johnson, M. J. Reagor, L. Frunzio, L. I. Glazman, S. M. Girvin, M. H. Devoret, and R. J. Schoelkopf, Phys. Rev. Lett. {\bf 107}, 240501 (2011).

\bibitem{Yamamoto2003} T. Yamamoto, Yu. A. Pashkin, O. Astafiev,
  Y. Nakamura, and J. S. Tsai, Nature {\bf 425}, 941 (2003).




\bibitem{Sillanpaa2011} A. Sillanp\"a\"a, J. I. Park, and R. W. Simmonds, Nature {\bf 449}, 438 (2007).

\bibitem{Majer2011} J. Majer, J. M. Chow, J. M. Gambetta, J. Koch, B. R. Johnson, J. A. Schreier, L. Frunzio, D. I. Schuster, A. A. Houck, A. Wallraff, A. Blais, M. H. Devoret, S. M. Girvin, and R. J. Schoelkopf, Nature {\bf 449}, 443 (2007).

\bibitem{Niskanen07}A. O. Niskanen, K. Harrabi, F. Yoshihara, Y. Nakamura, S. Lloyd, and J. S. Tsai, Science {\bf 316}, 723 (2007).

\bibitem{Plantenberg07}J. H. Plantenberg, P. C. de Groot, C. J. P. M. Harmans, and J. E. Mooij, Nature {\bf 447}, 836 (2007).



\bibitem{DiCarlo2009} L. DiCarlo, J. M. Chow, J. M. Gambetta,
  L. S. Bishop, B. R. Johnson, D. I. Schuster, J. Majer, A. Blais,
  L. Frunzio, S. M. Girvin, and R. J. Schoelkopf, Nature {\bf 460},
  240 (2009).
\bibitem{Chow2011}J. M. Chow, A. D. C\'orcoles, J. M. Gambetta, C. Rigetti, B. R. Johnson, J. A. Smolin, J. R. Rozen, G. A. Keefe, M. B. Rothwell, M. B. Ketchen, and M. Steffen, Phys. Rev. Lett. {\bf 107}, 080502 (2011).
\bibitem{Dewes2012} A. Dewes, F. R. Ong, V. Schmitt, R. Lauro, N. Boulant, P. Bertet, D. Vion, and D. Esteve, Phys. Rev. Lett. {\bf 108}, 057002 (2012).
\bibitem{Chow2012}J. M. Chow, J. M. Gambetta, A. D. C\'orcoles, S. T. Merkel, J. A. Smolin, C. Rigetti, S. Poletto, G. A. Keefe, M. B. Rothwell, J. R. Rozen, M. B. Ketchen, and
  M. Steffen, Phys. Rev. Lett. {\bf 109}, 060501 (2012). 


\bibitem{DiCarlo2010}L. DiCarlo, M. D. Reed, L. Sun, B. R. Johnson,
  J. M. Chow, J. M. Gambetta, L. Frunzio, S. M. Girvin, M. H. Devoret,
  and R. J. Schoelkopf, Nature {\bf 467}, 574 (2010); M. Neeley,
  R. C. Bialczak, M. Lenander, E. Lucero, M. Mariantoni,
  A. D. O'Connell, D. Sank, H. Wang, M. Weides, J. Wenner,
  Y. Yin, T. Yamamoto, A. N. Cleland, and J. M. Martinis, Nature {\bf 467}, 570 (2010).

\bibitem{Lu12} X.-Y. L\"{u}, S. Ashhab, W. Cui, R. Wu, and F. Nori,
  New J. Phys. \textbf{14}, 073041 (2012).


\bibitem{Rigetti2005} C. Rigetti, A. Blais, and M. Devoret, Phys. Rev. Lett. {\bf 94}, 240502 (2005).

\bibitem{Bertet2006}P. Bertet, C. J. P. M. Harmans, and J. E. Mooij, Phys. Rev. B {\bf 73}, 064512 (2006).

\bibitem{Liu06} Y.-X. Liu, L. F. Wei, J. S. Tsai, and F. Nori,
Phys. Rev. Lett. {\bf 96}, 067003 (2006).

\bibitem{Ashhab2006} S. Ashhab, S. Matsuo, N. Hatakenaka, and F. Nori, Phys. Rev. B {\bf 74}, 184504 (2006);
S. Ashhab and F. Nori, Phys. Rev. B {\bf 76}, 132513 (2007).


\bibitem{Rigetti2010}C. Rigetti and M. Devoret, Phys. Rev. B {\bf 81}, 134507 (2010). 




%\bibitem{Grajcar2005}M. Grajcar, A. Izmalkov, S. H. W. van der Ploeg,
%S. Linzen,E. Il'ichev, Th. Wagner, U. H\"ubner, H.-G. Meyer, Alec
%Maassen van den Brink, S. Uchaikin, and A. M. Zagoskin, Phys. Rev. B
%{\bf 72}, 020503 (2005). 

%\bibitem{Majer2005} J. B. Majer, F. G. Paauw, A. C. J. ter Haar, C. J. P. M. Harmans, and J. E. Mooij, Phys. Rev. Lett. {\bf 94}, 090501 (2005) (dc, away from optimal point).





\bibitem{Niskanen2006}A. O. Niskanen, Y. Nakamura, and J. S. Tsai,
  Phys. Rev. B {\bf 73}, 094506 (2006).
\bibitem{Harrabi2009}K. Harrabi, F. Yoshihara, A. O. Niskanen, Y. Nakamura, and J. S. Tsai, Phys. Rev. B {\bf 79}, 020507(R) (2009).

\bibitem{Grajcar2006} M. Grajcar, Y.-X. Liu, F. Nori, and A. M. Zagoskin,
Phys. Rev. B {\bf 74}, 172505 (2006).

%\bibitem{Plourde2004}B. L. T. Plourde, J. Zhang, K. B. Whaley, F. K. Wilhelm,
%T. L. Robertson, T. Hime, S. Linzen, P. A. Reichardt,
%C.-E. Wu, and J. Clarke, Phys. Rev. B {\bf 70}, 140501(R) (2004).

%\bibitem{Hime2006} T. Hime, P. A. Reichardt, B. L. T. Plourde,
%T. L. Robertson, C.-E. Wu, A. V. Ustinov, J. Clarke, Science {\bf 314}, 1427 (2006). 



%\bibitem{Van der Ploeg2007}S. H. W. van der Ploeg, A. Izmalkov, Alec
%Maassen van den Brink, U. H\"ubner, M. Grajcar, E. Il'ichev,
%H.-G. Meyer, and A. M. Zagoskin Phys. Rev. Lett. {\bf 98}, 057004 (2007). 

%\bibitem{Harris2007}R. Harris, A. J. Berkley, M. W. Johnson,
%P. Bunyk, S. Govorkov, M. C. Thom, S. Uchaikin, A. B. Wilson,
%J. Chung, E. Holtham, J. D. Biamonte, A. Yu. Smirnov, M. H. S. Amin,
%and Alec Maassen van den Brink, Phys. Rev. Lett. {\bf 98}, 177001
%(2007). 




\bibitem{Rabitz88}A. P. Peirce, M. A. Dahleh, and H. Rabitz, Phys. Rev. A {\bf 37}, 4950 (1988).
\bibitem{Tannor92} D. J. Tannor, V. A. Kazakov and V. Orlov,  in {\it Time-Dependent Quantum Molecular Dynamics},
  edited by J. Broeckhove and L. Lathouwers, NATO Advanced Studies Institute, Series B: Physics (Plenum Press, New
York, 1992), Vol. 299, pp. 347-360.
%\bibitem{Tannor93}J. Soml\'oi, V. A. Kazakov, and D. J. Tannor,
%Chem. Phys. 172, 85 (1993).

\bibitem{Kosloff02} J. P. Palao and R. Kosloff, Phys. Rev. Lett. \textbf{89}, 188301 (2002); J. P. Palao and R. Kosloff, Phys. Rev. A \textbf{68}, 062308 (2003).
\bibitem{Khaneja05} 
N. Khaneja, T. Reiss, C. Kehlet, T. Schulte-Herbr\"uggen, and S. J. Glaser, J. Magn. Reson. {\bf 172}, 296 (2005); 
%\bibitem{Sporl07} 	
%\bibitem{Tsai09}
D.-B. Tsai, P.-W. Chen, and H.-S. Goan, Phys. Rev. A
  \textbf{79}, 060306(R) (2009). 
%\bibitem{Brif10}C. Brif, R. Chakrabarti1 and H
\bibitem{Sporl07} A. Sp\"orl, T. Schulte-Herbr\"uggen, S. J. Glaser, V. Bergholm, M. J. Storcz, J. Ferber, and F. K. Wilhelm, Phys. Rev. A \textbf{75}, 012302 (2007).
\bibitem{Tannor11}R. Eitan, M. Mundt, and D. J. Tannor, Phys. Rev. A \textbf{83}, 053426 (2011).

\bibitem{Montangero2007} S. Montangero, T. Calarco, and R. Fazio, Phys. Rev. Lett. \textbf{99}, 170501 (2007).
\bibitem{Nielsen08} I. I. Maximov, Z. To\v{s}ner, and N. C. Nielsen, J. Chem. Phys. \textbf{128}, 184505 (2008).

%\bibitem{Jirari2006} H. Jirari and W. P\"otz, Phys. Rev. A,
%  \textbf{74}, 022306 (2006).
\bibitem{potz2008} M. Wenin and W. P\"otz, Phys. Rev. A \textbf{78},
  012358 (2008); M. Wenin and W. P\"otz, Phys. Rev. B \textbf{78},
  165118 (2008);  M. Wenin, R. Roloff, and W. P\"otz, J.
  Appl. Phys. \textbf{105}, 084504 (2009).
\bibitem{potz2009} R. Roloff and W. P\"otz, Phys. Rev. B \textbf{79},
  224516 (2009); M. Wenin and W. P\"otz, Phys. Rev. A \textbf{74},
  022319 (2006).
\bibitem{Jirari2009} H. Jirari, Europhys. Lett.
  \textbf{87}, 40003 (2009).

\bibitem{Rebentrost2009} P. Rebentrost, I. Serban, T. Schulte-Herbr\"uggen, and F. K. Wilhelm, Phys. Rev. Lett. \textbf{102}, 090401 (2009).

\bibitem{Hwang2012}B. Hwang and H.-S. Goan, Phys. Rev. A
  \textbf{85}, 032321 (2012).
\bibitem{Tai2014} J.-S. Tai, K.-T. Lin and H.-S. Goan, 
Phys. Rev. A \textbf{89}, 062310 (2014).

\bibitem{Aliferis2009} P. Aliferis and J. Preskill, Phys. Rev. A \textbf{79},
  012332 (2009).
\bibitem{Wang11} D. S. Wang, A. G. Fowler, and L. C. L. Hollenberg, Phys.
Rev. A \textbf{83}, 020302(R) (2011).
\bibitem{Fowler12L}A. G. Fowler, A. C. Whiteside, and L. C. L. Hollenberg,
Phys. Rev. Lett. \textbf{108}, 180501 (2012),

\bibitem{Fowler12}A. G. Fowler, M. Mariantoni, J. M. Martinis, A.
N. Cleland,  Phys. Rev. A \textbf{86}, 032324 (2012) 

\bibitem{Majer2005} J. B. Majer, F. G. Paauw, A. C. J. ter Haar, C. J. P. M. Harmans, and J. E. Mooij, Phys. Rev. Lett. {\bf 94}, 090501 (2005), and references therein.

\bibitem{Orlando1999}
T. P. Orlando, J. E. Mooij, L. Tian, C. H. van der Wal, L. S. Levitov,
S. Lloyd, and J. J. Mazo, Phys. Rev. B {\bf 60}, 15398 (1999).

\bibitem{Liu2005}Y.-X. Liu, J. Q. You, L. F. Wei, C. P. Sun, and F. Nori, Phys. Rev. Lett. {\bf 95}, 087001 (2005).
\bibitem{Deppe2008}F. Deppe, M. Mariantoni, E. P. Menzel, A. Marx, S. Saito, K. Kakuyanagi, H. Tanaka, T. Meno, K. Semba, H. Takayanagi, E. Solano, and R. Gross, Nat. Phys. {\bf 4}, 686 (2008). 


\bibitem{krotov1996} V. F. Krotov, {\it Global Methods in Optimal
    Control Theory} (Marcel Dekker, New York, 1996).


%\bibitem{Reich12}D. M. Reich, M. Ndong, and C. P. Koch, J. Chem.
%Phys. {\bf 136}, 104103 (2012).

\bibitem{Izmalkov2004}A. Izmalkov, M. Grajcar, E. Il'ichev, Th. Wagner, H.-G. Meyer,
A. Yu. Smirnov, M. H. S. Amin, A. Maassen van den Brink, and
A. M. Zagoskin, 
Phys. Rev. Lett. {\bf 93}, 037003 (2004).

\bibitem{You2005}J. Q. You, Y. Nakamura, and F. Nori, Phys. Rev. B {\bf 71}, 024532 (2005). 


\bibitem{Carmichael99}H. J. Carmichael, 
{\it Statistical Methods in Quantum Optics 1} (Springer, Berline, 1999).
\bibitem{Gardiner00}C. W. Gardiner and P. Zoller, {\it Quantum Noise}, 2nd edition. (Springer-Verlag, Berlin, 2000).
\bibitem{Breuer02} H.P.~Breuer and F.~Petruccione, {\it The Theory of Open Quantum Systems} (Oxford University Press, Oxford, 2002).



\bibitem{Bylander11} J. Bylander, S. Gustavsson, F. Yan, F. Yoshihara, K. Harrabi, G. Fitch, D. G. Cory, Y. Nakamura, J. S. Tsai, and W. D. Oliver, Nat. Phys. {\bf 7}, 565-570 (2011).



\bibitem{Egger14} D. J. Egger, and F. K. Wilhelm, arXiv:1402.7193. 
\bibitem{Magesan2012} E. Magesan, J. M. Gambetta, B. R. Johnson, C. A. Ryan,
J. M. Chow, S. T. Merkel, M. P. da Silva, G. A. Keefe,
M. B. Rothwell, T. A. Ohki, M. B. Ketchen, and M. Steffen,
Phys. Rev. Lett. {\bf 109}, 080505 (2012); 
E. Magesan, J. M. Gambetta, and J. Emerson, Phys.
Rev. Lett. {\bf 106}, 180504 (2011).
\bibitem{Chow2009}
J. M. Chow, J. M. Gambetta, L. Tornberg, J. Koch,
L. S. Bishop, A. A. Houck, B. R. Johnson, L. Frunzio,
S. M. Girvin, and R. J. Schoelkopf, Phys. Rev. Lett.
{\bf 102}, 090502 (2009);


\bibitem{Kelly14} J. Kelly, R. Barends et al.,  arXiv:1403.0035.

\bibitem{Nelder1967} J. A. Nelder and R. Mead, Comput. J. {\bf 7}, 308 (1967).
\bibitem{Rabitz1992} R. S. Judson and H. Rabitz, Phys. Rev. Lett. {\bf
    68}, 1500 (1992).
\bibitem{Rabitz2010} C. Brif, R. Chakrabarti1 and H. Rabitz, New J. Phys.
{\bf 12},  075008 (2010).

\bibitem{Dolde13} F. Dolde, V. Bergholm, Y. Wang, I. Jakobi,
  S. Pezzagna, J. Meijer, P. Neumann, T. Schulte-Herbr\"uggen
  J. Biamonte, and J. Wrachtrup, Nat. Commun. {\bf 5}, 3371 (2014).  

\end{thebibliography}
\end{document}